\documentclass{aa}
\usepackage[dvips]{graphicx,color}
\usepackage[]{txfonts}
\usepackage{natbib}
\bibpunct{(}{)}{;}{a}{}{,} % to follow the A&A style
\def\simless{\mathbin{\lower 3pt\hbox
{$\rlap{\raise 5pt\hbox{$\char'074$}}\mathchar"7218$}}}   %< or of order
\def\simmore{\mathbin{\lower 3pt\hbox
{$\rlap{\raise 5pt\hbox{$\char'076$}}\mathchar"7218$}}}   %> or of order
\newcommand{\be}{\begin{equation}}
\newcommand{\ee}{\end{equation}}

\newcommand{\dsfrac}[2]{\displaystyle{\frac{#1}{#2}}}

\begin{document}

\title{Deceleration of arbitrarily magnetized GRB ejecta:\\ the complete
  evolution}
\titlerunning{Deceleration of arbitrarily magnetized GRB ejecta} 
\authorrunning{Mimica, Giannios \& Aloy}
\author{P. Mimica\inst{1} \and D. Giannios\inst{2} \and
  M.~A. Aloy\inst{1}}

\institute{Departamento de Astronom\'ia y Astrof\'isica, Universidad
  de Valencia, 46100, Burjassot, Spain \and Max Planck Institute for
  Astrophysics, Box 1317, D-85741 Garching, Germany }

\offprints{Petar.Mimica@uv.es}
\date{Received / Accepted}

\abstract
% context
{The role of magnetic fields in gamma-ray burst (GRB) flows remains
  debated. If strong enough, they can leave their signature on the
  initial phases of the afterglow by substantially changing the
  backreaction of the flow as a consequence of its interaction with
  the external medium.}
% aims
{We aim to quantitatively understand the dynamical effect and
  observational signatures of magnetization of the GRB ejecta on the
  onset of the afterglow.}
% method
{We perform ultrahigh-resolution one-dimensional relativistic MHD
  simulations of the interaction of a radially expanding, magnetized
  ejecta with the interstellar medium. The need of ultrahigh numerical
  resolution derives from the extreme jump conditions in the region of
  interaction between the ejecta and the circumburst medium. We study
  the complete evolution of an ultrarelativistic shell all the way to
  a the self-similar asymptotic phase.}
% results
{Our simulations show that the complete evolution can be characterized
  in terms of two parameters, namely, the $\xi$ parameter introduced
  by Sari \& Piran (1995)
and the magnetization $\sigma_0$. We exploit this
  property by producing numerical models where the shell Lorentz
  factor is $\gamma_0\sim$ tens and rescaling the results to
  arbitrarily large $\gamma_0$. We find that the reverse shock is
  typically very weak or absent for ejecta characterized by
  $\sigma_0\simmore 1$.  The onset of the forward shock emission is
  strongly affected by the magnetization. On the other hand, the
  magnetic energy of the shell is transferred to the external medium on
  a short timescale ($\sim$several times the duration of the
  burst). The later forward shock emission does not contain
  information for the initial magnetization of the flow. The
  asymptotic evolution of strongly magnetized shells, after they have
  suffered a substantial deceleration, resembles that of hydrodynamic
  shells, i.e., they fully enter in the Blandford-McKee self-similar
  regime.}
{}

\maketitle

\section{Introduction} 
\label{intro}

It is generally believed that gamma ray bursts (GRBs) are the result
of the energy release in an ultrarelativistic outflow. The mechanisms
responsible for launching, accelerating the flow and powering the GRB
emission are still not well understood. Two alternatives for the
energy content of the flow have been explored. The GRB flow may be
initially dominated by thermal energy density creating a
  fireball (\citealt{Goodman:1986eq,Paczynski:1986qd}) or by the
energy stored in magnetic fields giving rise to a Poynting-flux
dominated flow (PDF ;
\citealt{Usov:1992hp,Thompson:1994ft,Meszaros:1997bx}).

In fireball models, magnetic fields are not dynamically important at
any stage of the flow evolution. On the other hand, models of MHD jet
acceleration indicate that the conversion of Poynting flux to kinetic
energy is partial \citep{Michel:1969kx,Li:1992ez,Beskin:1998vn,
  Vlahakis:2003bl,Giannios:2006uk}.  As a result, the flow remains
rather strongly magnetized at large distance where it interacts with
the circumburst medium.

The interaction of the fast flow with the external medium likely
powers the afterglow emission. The initial phases of this interaction
depend, to a large extent, on the magnetization of the
flow. Strong magnetic fields affect the shock conditions 
and the internal dynamics of the ejecta.
\citet{Rees:1974yq} and \citet{Kennel:1984kx} have paved the way to
understanding the effect of the magnetization on the shock
conditions. They showed that, in the ideal MHD limit, shocks in
magnetically dominated flows cannot efficiently dissipate energy. This particular effect has
been studied recently by examining the shock conditions of the
(reverse) shock that propagates into the GRB flow and the resulting
emission \citep{Fan:2004if,Zhang:2005ts}. Taking the internal
evolution of the magnetized shell into account, \cite{Giannios:2008zl}
have argued that even moderately strong magnetic fields can suppress
the reverse shock altogether possibly explaining the observed paucity
of reverse shock signature in the early afterglow.
 
Here, we complete previous (semi-)analytical studies of the
afterglow phase of magnetized ejecta with relativistic MHD
simulations. We follow the deceleration of magnetized ejecta from the
initial phases of the interaction up to the self similar regime where
all the energy has been transferred to the shocked external
medium. These simulations clarify what are the dynamical effects of
magnetization of the GRB ejecta and their observational implications
connected to the forward and reverse shock emission.  Along the
  way, we will show a new set of scaling laws \S\ref{sec:rescaling}
  that enable us to extrapolate the results of numerical models with
  moderate values of the initial bulk Lorentz factor ($\sim 15$) of
  the ejecta to {\it equivalent} models with much larger Lorentz
  factors $\simmore 100$.

\section{Ejecta-medium interaction}
\label{interaction}

At large distances from the central engine, there is a substantial
interaction of the relativistic ejecta with the external medium. This
interaction is believed to result in the afterglow emission. An
important difference between fireballs and PDFs lies in the
magnetization of the ejecta at the onset of the afterglow phase.  In
fireball models, the energy of the flow is dominated by the kinetic
energy of baryons at large distance from the central engine.  If the
flow is launched Poynting-flux dominated, it is expected to maintain a
large fraction of its energy in the form of magnetic energy, the rest
being in kinetic form
\citep{Michel:1969kx,Li:1992ez,Beskin:1998vn,Drenkhahn:2002zm,Lyutikov:2003lk}. Since
(at least the initial phases) of the ejecta-external medium
interaction depend on the magnetization of the ejecta
  \citep{Kennel:1984kx,Fan:2004if,Zhang:2005ts,Genet:2007cl}, it is
possible to discriminate among fireballs and PDF models from afterglow
observations.  As we show in this work, early afterglow observations
are particularly promising in this respect.

\subsection{Previous Studies}

The deceleration of {\it non-magnetized} ejecta has been well studied
with both analytical \citep{Rees:1992kx,Sari:1995oq} and with
numerical approaches in one dimension (1D) \citep{Kobayashi:1999ly}
and two dimensions (2D) \citep{Granot:2001fq,Meliani:2007qm}. The 2D
studies are important to follow the late-time lateral spreading of
collimated ejecta \citep{Rhoads:1999rt}. On the other hand, the
initial phases of the deceleration of the ejecta, in which we are
interested here, are not affected by 2D effects and can be studied
assuming spherically symmetric flow.

\citet{Sari:1995oq} considered the case of non-magnetized ejecta
assuming a cold shell with (isotropic equivalent) kinetic energy $E$,
Lorentz factor $\gamma_0$ and width $\Delta_0$, which
moves against external medium with density $\rho_{e}$. This
interaction leads to a pair of shocks: one that propagates in the
external medium (forward shock) and one that slows down the ejecta
(reverse shock).  The strength of the reverse shock depends on the
ratio of the densities of the shell and the external medium and on the
bulk Lorentz factor of the flow. It can be shown that the strength of
the reverse shock can be conveniently parametrized by
\be \xi\equiv \sqrt{\frac{l}{\Delta_0}}\frac{1}{\gamma_0^{4/3}},
\label{ksi}
\ee 
where $l=(3E/4\pi n_{\rm e} m_{\rm p}c^2)^{1/3}$ is the Sedov length,
$n_{\rm e}$ is the number density of electrons in the external medium
and $m_{\rm p}$ is the proton mass. $E$ is the total energy
(kinetic in this case) of the ejecta. In the limit where $\xi\gg 1$ the
reverse shock is Newtonian and the shell is said to be
``thin''. The ejecta do not decelerate much by the time the reverse
shock crosses them. If $\xi\ll 1$, we find ourselves in the ``thick
shell'' case, and the reverse shock is relativistic and slows down the
ejecta appreciably \citep{Sari:1995oq}. . For typical parameters of
GRB flows $\xi$ is of order of unity with $0.1\simless \xi\simless$
several.

The dynamics of the deceleration of strongly magnetized ejecta have
not been studied in the same detail. In addition to the $\xi$
parameter, the ejecta are characterized by the magnetization
$\sigma_0$ defined as the ratio of magnetic-to-kinetic energy in the
flow.  \citet{Kennel:1984kx} solved for the ideal MHD shock conditions
for arbitrarily magnetized ejecta with a dominant toroidal field, and
showed that the dissipation by the shock gets weaker as $\sigma_0$
increases \citep[e.g][]{Lyutikov:2003lk}. They applied their analysis
to the standing shock of pulsar winds. More recently
\citet{Zhang:2005ts} focused on the effect of magnetization in the
context of GRB afterglows.  They ignored the internal evolution of the
shell prior to the interaction with the external medium and studied
the reverse shock crossing phase (provided that there is a reverse
shock forming). They found distinct features in the early time
light curves because of the magnetization. The shocks from the
interaction of the GRB ejecta with the external medium propagate
forwards and the shock conditions depend on the distance from the
central engine.  The \citet{Zhang:2005ts} analysis has been criticized
by \citet{Lyutikov:2005tx} for the assumption on the distance where
the ejecta decelerate.  \citet{Giannios:2008zl} took into account the
internal evolution of the ejecta and derived the an analytic
condition for existence of a reverse shock depending on $\xi$ and
$\sigma_0$ in a parameter space relevant for GRB flows.\footnote{ At
  the time of the referring of this work, \citet{Mizuno:2008kx} have
  also published a work addressing the problem of the deceleration of
  arbitrary magnetized ejecta into an unmagnetized medium, and discuss
  its implications for GRBs and active galactic nuclei. However, in
  spite of the undoubtable academic interest of their studies, the
  conditions set by these authors (particularly, the use of planar
  symmetry, and the very small density contrast between the magnetized
  shell and the external medium $\sim 100$) are far from those met in
  GRB afterglows, specially during the early afterglow propagation.}

After the reverse shock (if there is one) reaches the back part of the
ejecta, there is a transient phase of interaction where rarefaction
waves cross the shocked ejecta and shocked external medium. Gradually
most of the energy is passed in the shocked external medium and the
whole structure relaxes to the self-similar blast wave described in
\citet{Blandford:1976yg}. From this point on, the evolution of the
blast wave depends only on the total energy $E$ and the density of the
external medium $n_{\rm e}$ and {\it not} on $\sigma_0$. After the
self-similar evolution has been reached, nothing can be inferred about
the initial magnetization of the flow.

However, none of these studies have addressed two important
questions. First, there is a question in which stage of the
interaction a reverse shock forms (if it forms at all).  At short
distance from the central engine the magnetic pressure of the shell is
high enough that the shell rarefies upon interacting with the external
medium. This rarefaction may turn into a (reverse) shock at larger
distance where the magnetic pressure in the shell drops.  Second,
although it is clear that (ignoring radiative losses) the total energy
initially in the shell is passed onto the external medium at a
distance of the order of the Sedov length which is independent of the
magnetization of the flow \citep{Lyutikov:2005tx}, the details of how
exactly this happens depend on the magnetization.  These two aspects
are closely connected to the energy that is dissipated in the (forward
and reverse) shocks as function of distance and, consequently,
to the afterglow emission from particles accelerated in these shocks.
We address this issue here by studying the full dynamical interaction
from the initial stages all the way to the establishment of the
self-similar evolution.  To this end we perform ultra-high resolution
one-dimensional relativistic MHD simulations of shell-medium
interaction.

\section{The model for the ejecta}

We focus on the GRB flow at a distance where there is substantial
interaction with the external medium. This interaction likely takes
place well after the acceleration, collimation and prompt emission
phases are over.  After the internal dissipation phase (believed to
power the $\gamma$-ray emission) finishes, the flow expands radially
and cools down. The expansion also leads to a dominant toroidal
component for the magnetic field. At the, so-called, Alfv\'en point,
the poloidal $B_{\rm p}$ and toroidal $B_{\phi}$ field components are
expected to be of similar magnitude. Further out, the flux freezing
condition results in $B_{\rm p}\propto 1/r^2$ while the induction
equation predicts slower decline for $B_{\phi}\propto 1/r$. The same
scalings hold if the initial $B$-field were random resulting in
$B_{\phi}\gg B_{\rm p}$ at large distance from the central engine.

We consider radially moving, cold shell of ejecta of width $\Delta_0$,
total (kinetic and magnetic) energy $E$ that coasts with a bulk
Lorentz factor $\gamma_0$. The magnetic content of the flow is
parametrized with the magnetization parameter $\sigma_0$ which stands
for the magnetic-to-kinetic energy ratio in the shell.  The flow is
assumed to move with super-fast magnetosonic speeds (i.e.,
$\gamma_0^2>1+\sigma_0$; for studies of the opposite limit see
\citealp{Lyutikov:2006wj,Genet:2007cl}). For the simulations
presented bellow, the shell is located at an initial distance $r_0$
from the central engine. The choice of $r_0$ is important since $r_0$
should be small enough not to affect the subsequent interaction of the
ejecta with the external medium.  $r_0$ must be set smaller than any
of the characteristic radii that appear when considering the
deceleration of magnetized ejecta.  These radii are the `contact'
radius and the `reverse shock crossing' radius to be defined in the
next section.

\subsection{Characteristic distances}

In the super-fast magnetosonic flow under consideration, the various
parts along the radial direction have dropped out of MHD contact
during the acceleration phase. It can be shown that for a cold flow
that coasts with constant speed with dominant toroidal field the
magnetization remains constant. The time it takes for a fast MHD wave
to cross the width of the shell is therefore fixed. The expansion
timescale $t_{\rm exp}=r/\gamma_0c$ is initially much shorter than
that of MHD waves but increases linearly with distance from the
central engine. At the so-called `contact' radius $r_{\rm c}$ MHD
waves cross the width of the shell on a timescale comparable to the
expansion timescale \citep{Giannios:2008zl}
\be r_{\rm c}\simeq
\Delta_0\gamma_0^2\Big(\sqrt{\frac{1+\sigma_0}{\sigma_0}}-1\Big) .
\label{rc}
\ee
After contact is established, the shell is no longer in pure ballistic
motion and internal evolution because of MHD forces can no longer be
ignored.  On the other hand in non-magnetized ejecta the sound speed
drops fast with distance because of adiabatic expansion, and the
motion is is not affected by the pressure of the shell.

A second important radius is the radius where the reverse shock
reaches the rear part of the ejecta. This radius is derived by
\citet{Zhang:2005ts} by solving the ideal MHD shock conditions for
arbitrarily magnetized ejecta (see also \citealt{Fan:2004if} for the
case of mildly magnetized ejecta).  Their analysis describes the
reverse shock crossing phase provided that there is a reverse shock
forming. The reverse shock crossing radius can be approximately
expressed as \citep{Giannios:2008zl}
\be \label{eq:rrs} r_{{\rm rs}} \simeq
l^{3/4}\Delta_0^{1/4}/\sqrt{1+\sigma_0}.
\label{rsB}
\ee 
The \citeauthor{Zhang:2005ts} analysis does not take into account the
internal evolution of the shell. It is thus accurate when such
evolution is not significant, i.e. when $r_{\rm rs}<r_{\rm c}$.

The initial distance $r_0$ where the shell is set up must be $r_0\ll
{\rm min} [r_{\rm c}, r_{\rm rs}]$ so that the simulation starts early
enough to follow both any rarefaction waves within the shell, and
shock waves result from the interaction with the external medium.

\subsection{Characteristic quantities}

In this paper we frequently use the following definition
of the normalized time of observation for a model with parameters
$\gamma_0$, $\Delta_0$ and $r_0$:
\be\label{eq:tobs}
    t_{\rm obs} := \Delta_0^{-1} \left[ct - r\right]\, , 
\ee
where $t_{\rm obs}$ is the time of observation of a signal sent from
radius $r$ at time $t$ in the GRB frame or laboratory frame,
normalized to the light crossing time of the initial width of the
shell $\Delta_0/c$. As we will show in Sec.~\ref{sec:invariance}, this
definition of $t_{\rm obs}$ enables us to compare properties (in the
observer frame) of shells with the same $\xi$ independent of their
initial Lorentz factor.

We also often base our arguments on the relative Lorentz factor
$\gamma_{\rm rel}$ between two parts of the fluid separated by the
shock front. For ultrarelativistic flows we use
\begin{equation}\label{eq:gamma_rel}
  \gamma_{\rm rel}:= \dsfrac{1}{2}\left(\frac{\gamma_a}{\gamma_b} +
    \frac{\gamma_b}{\gamma_a} \right)\, ,
\end{equation}
where $\gamma_a$ and $\gamma_b$ are the Lorentz factors of the fluid
ahead and behind of the shock, respectively. We point out that
$\gamma_{\rm rel}$ depends only on the ratio $\gamma_a/\gamma_b$.

\subsection{Numerical models}

Although the problem is characterized by several parameters $E$,
$n_{\rm e}$, $\Delta_0$, $\gamma_0$ and $\sigma_0$, it turns out that
for the systematic study of the shell-medium interaction we need to
focus on the combination of the first four ones
parametrized by

\be 
\xi\propto  (E/n_{\rm
  e})^{1/6}/\Delta_0^{1/2} \gamma_0^{4/3}
\ee
\noindent
and $\sigma_0$. We demonstrate and quantify this statement in the next
Section. In order to simplify the analysis, we restrict ourselves to
the case in which the external medium density is uniform ($\rho_{\rm
  e} \approx 3\times 10^{-4}\rho_0 (1+\sigma_0)$; $\rho_0$ being the
initial shell density, and leave the study of stratified external
media for a future work.

As in the case of unmagnetized ejecta, we make use of the Sari-Piran
parameter $\xi$ in order to partly characterize the strength of the
reverse shock. Certainly, in the magnetized case, the shock strength
is not uniquely set by $\xi$. Instead, an additional parameter,
$\sigma_0$, needs to be introduced to fully describe the reverse shock
strength of arbitrarily magnetized flows. Thus, one deals with a
$\xi-\sigma_0$ plane in exploring different cases for the initial
phases of shell-external medium interaction. Here, we explore the $\xi
\sim 1$ regime that is relevant for typical GRB parameters. Numerical
reasons limit us to the $0\le \sigma_0\le 3$ range for the
magnetization parameter.

Our model runs are summarized in Tab.~[\ref{table:1}]. The $\xi=1.1$
runs (thin shells) are characterized by $E=3.33\times 10^{53}$ erg,
$\Delta_0=10^{15}$ cm, $\gamma_0=15$, $n_{\rm e}=10$ cm$^{-3}$. The
$\xi=0.5$ models (thick shells) have ten times larger total energy $E$
and width of the shell $\Delta_0$ while $\gamma_0$ and $n_{\rm e}$
remain fixed.  The ``continuous flow'' model (to be discussed in more
detail in Sect. 4.3) describes a flow of constant total
(kinetic+Poynting) luminosity of $L=10^{49}$ erg/sec that moves with
$\gamma_0=15$ and collides with external medium of number density
$n_{\rm e}=10$ cm$^{-3}$. In all models $r_0 = 5\times 10^{16}$ cm.
\begin{table}
{\large
\caption{Parameters of the numerical models}
\label{table:1}
\centering 
\begin{tabular}{|l|c|c|c|}
  \hline
  $\sigma_0$ & 0 &1 & 3\\
  \hline
  thin shell ($\xi = 1.1$) & $\surd$ & $\surd$ & \\
  \hline
  thick shell ($\xi = 0.5$) & $\surd$ & $\surd$ & \\
  \hline
  continuous flow & & $\surd$ & $\surd$\\
  \hline
\end{tabular}
}
\end{table}

One may notice that the model runs are characterized by
unrealistically low Lorentz factor $\gamma_0=15$ and wide shells
$\Delta_0\sim 10^{15}$ cm with respect to what is expected from a GRB
flow (i.e. $\gamma_0\simmore 100$, $\Delta_0\simless 3\times 10^{12}$
cm). This choice of parameters is made so that the problem is reliably
resolved with our RMHD code. While runs with $\gamma_0\simmore 100$ in
combination with extreme density and magnetic field jumps at the edge
of the ejecta shell are not feasible at this stage, we propose a
method to extrapolate the results of the $\gamma_0=15$ simulation by
appropriately rescaling of the initial conditions. Furthermore, we
have run simulations with $\gamma_0=10, 20$ where we demonstrate the
accuracy of the rescaling procedure (see Section 4.4).

\section{Results}
\label{results}

To derive the results presented in this Section, we solve the
equations of RMHD in 1D spherical geometry with magnetic field
perpendicular to the direction of propagation of the fluid, i.e., with
a purely toroidal magnetic field. The system of RMHD equations, and
the numerical tests we have made to choose the appropriate numerical
resolution for our experiments (between $10^4$ and
  $6\times10^4$ cells to resolve the initial radial width of the
  ejecta) are shown in the Appendix~\ref{method}. Finally, all our
models have been run until the bulk Lorentz factor behind the forward
shock has dropped to $\gamma \sim 2-3$. By that time the shell has
suffered a substantial deceleration and fully entered in the
Blandford-MacKee self-similar regime.
\begin{figure}
\centering \includegraphics[scale=0.32]{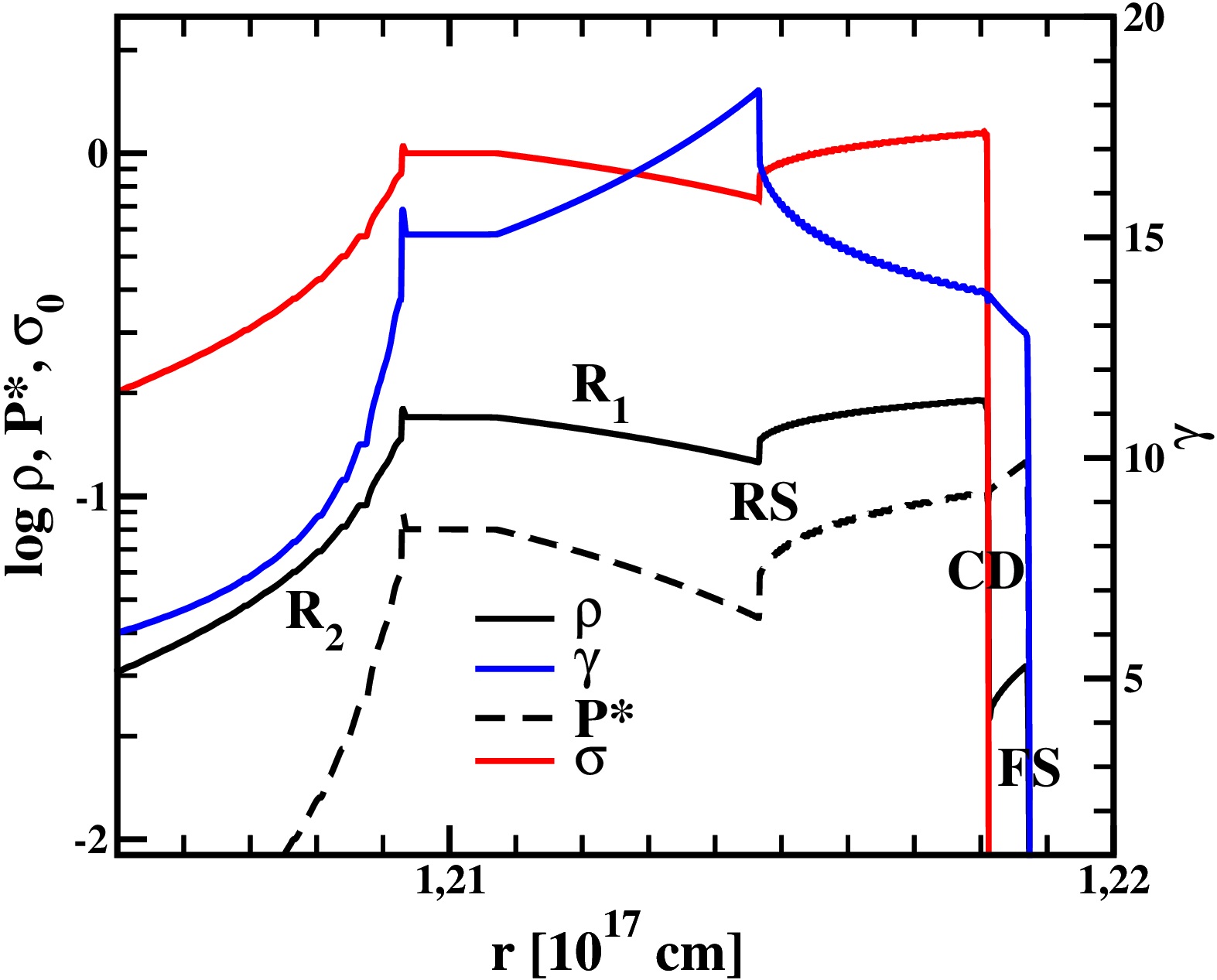}
\caption[]{Snapshot of the thin magnetized shell evolution taken after
  the RS has formed and before it has crossed the shell. Full and
  dashed black lines show the logarithms of the rest-mass density
  (normalized to the initial shell density $\rho_0$) and of the
  pressure (normalized to $\rho_0 c^2$). The red line shows the
  logarithm of the magnetization $\sigma$, while the blue line shows
  the fluid Lorentz factor $\gamma$ in the linear scale. All
  quantities are shown as a function of radius $r$. Positions of the
  forward shock ($FS$), contact discontinuity ($CD$), reverse shock
  ($RS$), and left- ($R_1$) and right-going ($R_2$) rarefactions have
  been indicated. There exist another rarefaction (moving
  backwards in the external medium) and another contact discontinuity
  that form at the rear edge of the shell, which are not shown
  here. Both structures are located to the right of $R_2$. The
  unshocked external medium is located in front of the FS and its
  density $\rho_{\rm e} \approx 6\times 10^{-4}\rho_0$ is smaller than
  the minimum density shown in the plot. The rarefaction $R_2$, the CD
  and the FS display a qualitatively similar profile in the
  non-magnetized case. The rarefaction $R_1$ and the late steepening
  of the conditions at its tail, resulting into the formation of RS,
  are specific of magnetized ejecta.}
\label{fig:RScross}
\end{figure}

\subsection{Non-magnetized shells}

The $\sigma_0=0$ models show the well known features expected from
analytical results \citep{Sari:1995oq} and simulations
\citep{Kobayashi:1999ly}.  The {\it thin shell} ($\xi=1.1$) model is
characterized by a Newtonian-to-mildly-relativistic reverse shock.
The reverse shock crosses the shell at a distance $r_{\rm rs}=3\times
10^{17}$ cm which agrees within $\sim 10$\% with the analytical
estimate from Eq.~(\ref{eq:rrs}). At this distance, the relative
Lorentz factor of the shocked ejecta with respect to the unshocked
shell is $\gamma_{rel}\simeq 1.18$.  The thick shell ($\xi=0.5$) model
finds itself closer to the ``relativistic reverse shock'' regime with
$r_{\rm rs}=9.5\times 10^{17}$ cm (within $\sim 10$\% of the initial
estimate) and $\gamma_{rel}\simeq 1.60$. In both runs, after the
reverse shock crosses the shell, there is a rarefaction that starts at
the rear part of the shell and propagates forwards (a similar
rarefaction happens in the magnetized case, which we label $R_2$ in
Fig.~\ref{fig:RScross}). The rarefaction crosses the contact
discontinuity, generated in the leading radial edge of the shell (an
equivalent contact discontinuity arises in the magnetized case, see
``CD'' in Fig.~\ref{fig:RScross}), and reaches the forward shock when
the shell reaches $r\sim 1.6 r_{\rm rs}\simeq 5 \times 10^{17}$ cm and
$r\sim 1.6 r_{\rm rs}\simeq 1.6 \times 10^{18}$ cm, in the thin- and
in the thick-shell case, respectively. At this stage, $\sim 90$\% of
the energy of the shell has been transferred to the shocked external
medium.  Within a factor of $\sim 2$ in radius the blastwave fully
relaxes to the Blandford-McKee self-similar solution.

\begin{figure}
\centering \includegraphics[scale=0.32]{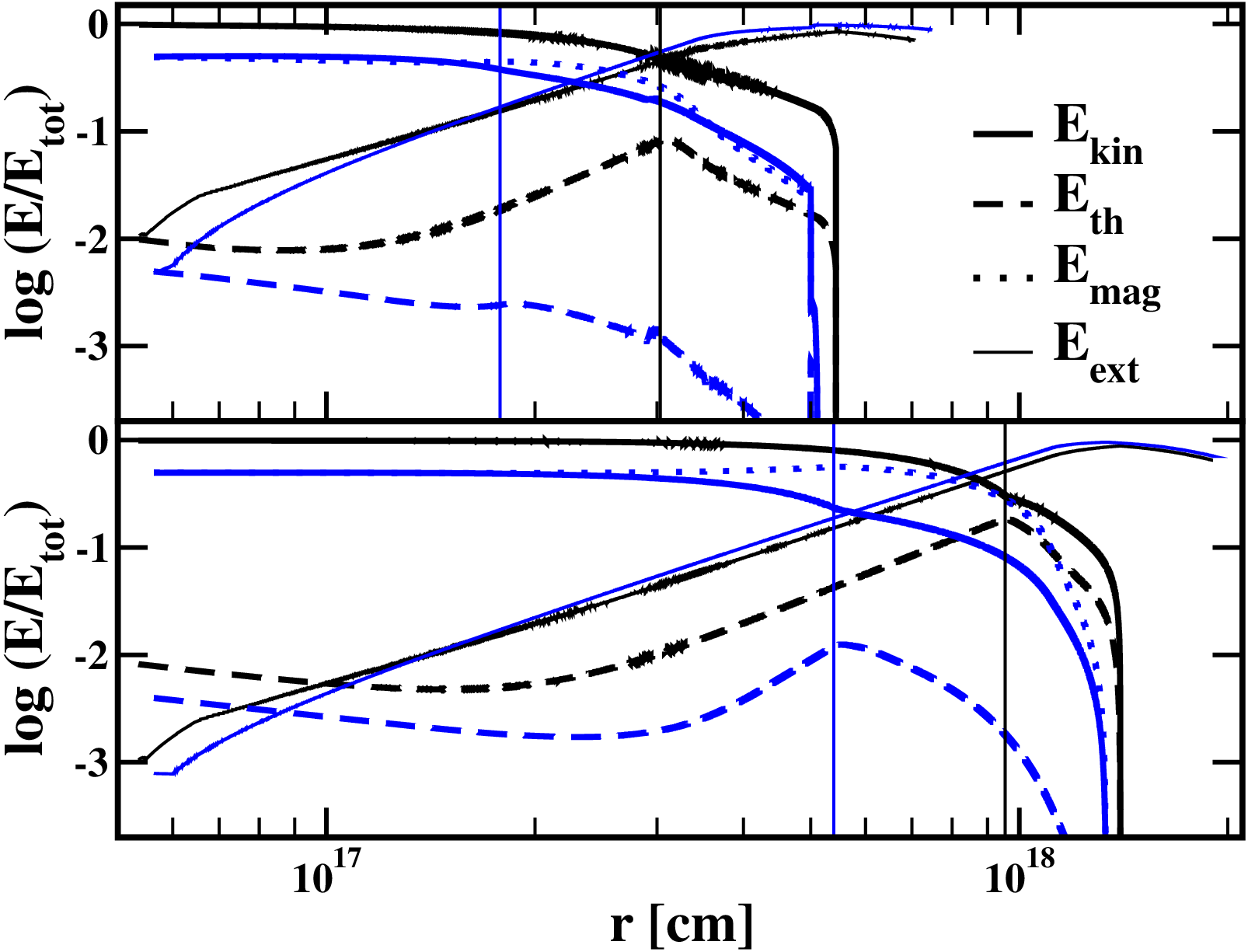}
\caption[]{ Energy in different components (normalized to
  the initial total energy in the shell) as a function of the radius
  of the FS. Upper and lower panels show the thin ($\xi =
  1.1$) and thick ($\xi = 0.5$) shell models, respectively. Black and
  blue lines correspond to un-magnetized ($\sigma = 0$) and
  magnetized ($\sigma = 1$) models, respectively. Thick full, dashed
  and dotted lines respectively show the kinetic, thermal and magnetic
  energy of the shell. The thin full line shows the total energy in
  the shocked external medium. Thick vertical lines denote the radius
  of the FS at the moment when the RS has crossed the shell.}
\label{fig:energy}
\end{figure}

The different components of the total energy \citep[see
e.g.,][]{Mimica:2007db} as function of radius of the front shock (FS)
are shown in Fig.~\ref{fig:energy}. For the thin shell model, the peak
of the thermal energy of the shell (approximately 9\% of the total
energy contained in the ejecta) traces the reverse shock (RS) crossing
of the shell. Beyond $5\times 10^{17}$ cm most of the energy that was
initially in the shell has been transferred to the shocked external
medium. The apparent ``disappearance'' of the shell at $r\approx
5.4\times 10^{17}\,$cm is a numerical artifact of the grid
re-mapping\footnote{As described by \citet{Mimica:2007db}, grid
  re-mapping enables us to follow the evolution of a localized shell
  over large distance by repeated re-mapping of the numerical grid. In
  this work the grid always follows the front shock, so that, once
  the shell has been slowed down by the reverse shock, it is
  eventually ``lost'' from the grid.}. However, this effect is
irrelevant for the discussion of the features we are interested in,
since all of them happen before the shell ``disappears'', both in
non-magnetized and magnetized models. For the unmagnetized thick shell
runs we see that the reverse shock dissipates more energy from the
shell, reaching approximately 18\% of the total shell energy by the
time it crosses the shell.

\begin{figure}
\centering \includegraphics[scale=0.32]{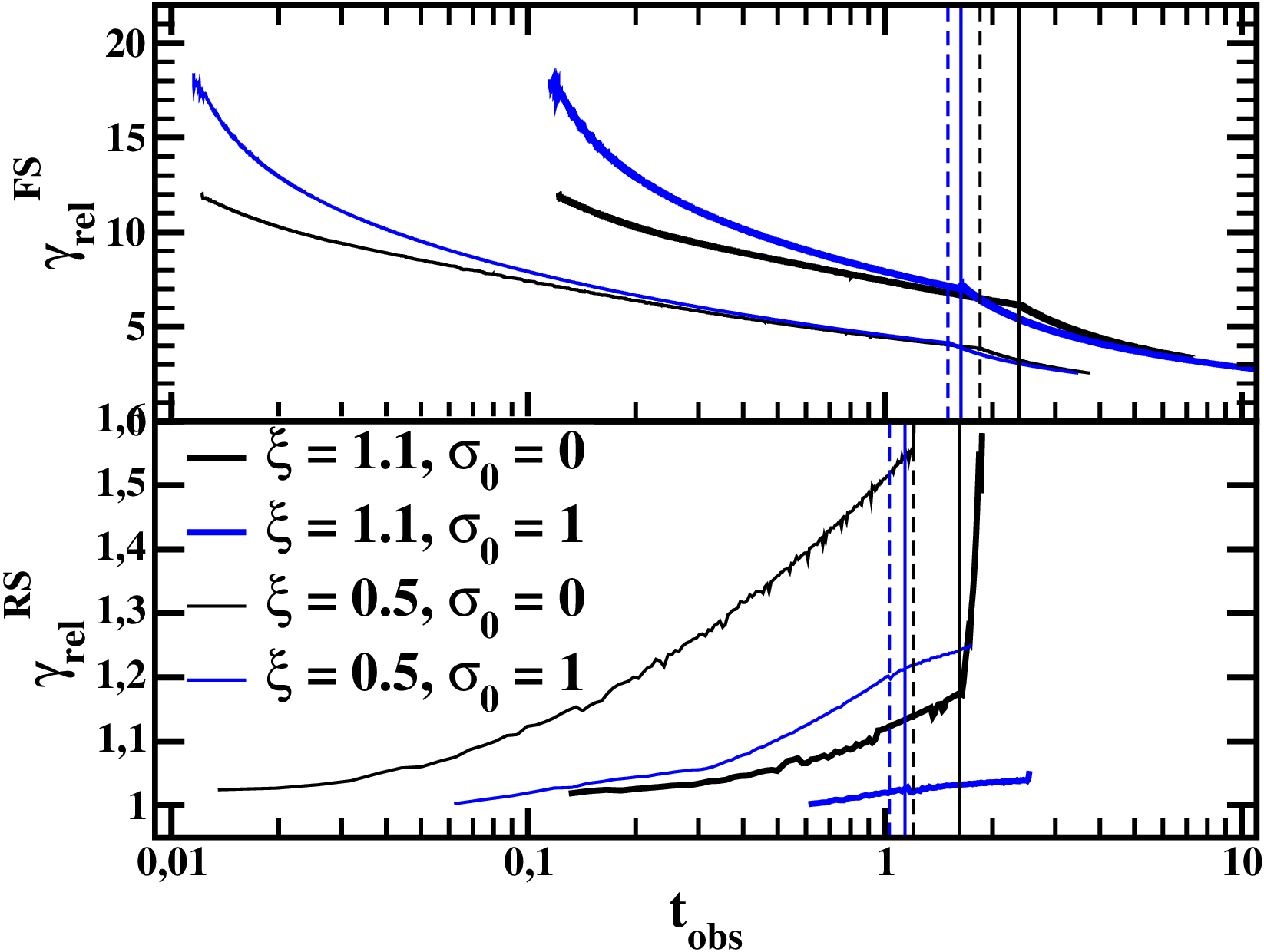}
\caption[]{Relative Lorentz factor at the FS (upper panel) and RS
  (lower panel) as a function of the normalized observer time $t_{\rm
    obs}$ (Eq.\ref{eq:tobs}). Full and dashed lines correspond to the
  thin and thick shell models, respectively. Black and blue colors
  denote non-magnetized and magnetized models, respectively. The
  vertical lines in the lower panels denote the time of observation
  when the RS crosses the shell. In the upper panel the vertical lines
  denote the time of observation when the rarefaction $R_2$, that
  originates from the rear edge of the shell and interacts with the
  RS, catches up with the FS.}
\label{fig:gammas}
\end{figure}

In Fig.~\ref{fig:gammas} we show the relative Lorentz factor at the FS
and the RS as function of observer time. Both shocks form immediately
after one lets the initial shell to evolve. The RS becomes stronger
with distance, as can be seen from the increase of $\gamma_{rel}$ with
$t_{\rm obs}$. This strengthening continues until it reaches the rear
part of the shell, where it encounters a much less dense medium, which
results in a kink in the RS curves (marked with vertical black lines
in Fig.~\ref{fig:gammas}). The peak of the emission associated with
the RS is expected to occur at the moment the RS breaks out of the
shell, since afterwards the density, pressure and velocity of the
shocked medium decrease abruptly, and precisely there it reaches its
maximum strength. For the thin shell case this happens slightly after
the burst, i.e. at observer time $t_{ \rm obs}\sim 1.6$. For the thick
shell the peak of the RS emission appears at the end of the burst, by
a time $t_{ \rm obs}\sim 1.2$.

The Lorentz factor of the external medium just behind the FS drops as
function of radius. An interesting feature is the change in the slope
of $\gamma_{\rm rel}^{\rm FS}(t_{\rm obs})$ at time $t_{\rm obs} =
2.5$ for the thin shell (at time $t_{\rm obs} = 1.84$ for the thick
shell) which is a result of the rarefaction $R_2$
(Fig.~\ref{fig:RScross}) reaching the forward shock. For a short
transient period the FS reduces its strength because the density
behind it is reduced by the action of the rarefaction $R_2$. The decay
of $\gamma_{\rm rel}^{\rm FS}(t_{\rm obs})$ is faster for $\Delta
t_{\rm obs}\sim 0.5$ after the rarefaction $R_2$ catches up with the
FS. Thus, we expect that the early afterglow emission weakens
transiently faster with time. Later, the time dependence of
$\gamma_{\rm rel}^{\rm FS}\propto t_{\rm obs}^{-3/8}$ expected from
the self-similar solution is gradually established.

\subsection{Magnetized shells}
\label{sec:magnetized_shells}

The initial phases of deceleration of strongly magnetized ejecta show
distinct difference with respect to unmagnetized ones related to the
magnetic pressure of the ejecta. As we discussed in
\citet{Giannios:2008zl}, the jump in the physical conditions existent
between the shell and the external medium results into the development
of two Riemann structures, one at every radial (rear and forward) edge
of the shell. The waves into which each of these Riemann structures
decompose are combinations of shocks and/or rarefactions separated by
contact discontinuities. If the magnetic pressure is sufficiently
large, instead of a typical double shock structure, a right-going
shock ($FS$) and a left going rarefaction ($R_1$) develop from the
forward edge of the shell (Fig.~\ref{fig:RScross}). Similarly, at the
backward edge of the shell, a rarefaction (not shown in
Fig.~\ref{fig:RScross}) moving backwards (in the shell comoving frame)
into the external medium develops, as well as a rarefaction develops
penetrating the shell.

The expansion of the shell leads to a decrease of the magnetic
pressure and the Lorentz factor of contact until the
``transition'' radius, $r_{\rm tr}$, is reached where
$\gamma_{CD}=\gamma_0$. This distance can be estimated by balancing
the pressure of the shocked external medium $P=4\gamma_0^2n_em_pc^2/3$
with the magnetic pressure of the shell $P_{\rm B}=B^2/\gamma_0^2
8\pi=E\sigma_0/8\pi r^2\gamma_0^2\Delta_0(1+\sigma_0)$ and solving for
the radius
\be
r_{\rm tr}=\Big(\frac{3E\sigma_0}{32\pi n_{\rm e} m_{\rm
    p}c^2\gamma_0^4\Delta_0 (1+\sigma_0)}\Big)^{1/2}.
\ee

From this distance onwards the shell slows down with respect to its
initial $\gamma_0$. Some time after the rarefaction $R_1$ has
propagated backwards into the shell, a new shock develops inside of
the rarefaction fan due to the radial expansion of the shell (RS in
Fig.~\ref{fig:RScross}). This shock sweeps backwards through the shell
and, therefore, it is effectively a {\it reverse shock}. Remarkably,
this shock does not immediately originate from the initial jump at the
leading radial edge of the shell. Instead, it develops at the faster
parts of the rarefaction fan and {\it not} directly at the contact
discontinuity separating the FS and $R_1$ (see
Fig.~\ref{fig:RScross}). The reason is that $\sigma \propto r^2\rho$
in the cold, magnetized shell (see Appendix A), and therefore it
decreases in the rarefaction fan. The formation of this shock can only
happen when the initial shell expands spherically, and not if the
shell is assumed to evolve under conditions of planar (Cartesian)
symmetry (as assumed in \citealt{Mizuno:2008kx}). The fact that
$\sigma$ decreases in the rarefaction, combined with the steepening of
the rarefaction profile due to the spherical geometry, leads to a
decrease of the fast magnetosonic speed in the whole rarefaction
fan. The decrease is larger right at the head of the rarefaction
where, eventually, a shock forms\footnote{ We thank the referee for
  pointing out the fact that, without this decrease of the fast speed,
  a shock would either form immediately or never.}.  We therefore
realize that the structure of the flow is much more complex than in
the non-magnetized case, since it has developed a RS inside of the
rarefaction fan of $R_1$.

The shock is initially weak ($\gamma_{\rm
  rel}^{\rm RS} \simeq 1$) and remains so during the period in which
it sweeps the whole (thin) shell (Fig.~\ref{fig:gammas}). When it
reaches the back edge of the shell $\gamma_{rel}\simeq 1.03$, i.e., it
remains still Newtonian. The reverse shock dissipates a negligible
amount of energy (some $\sim$0.1\% of the total energy in the shell).
It becomes stronger in case a thick shell is considered, reaching
$\gamma_{rel}\simeq 1.2$ by the time it reaches the rear radial edge
of the shell (Fig.~\ref{fig:gammas}, lower panel). Thus, these RS
sweeping a thick shell leads to a higher dissipation of energy ($\sim$
1\% of the total). However, the dissipated energy is still a factor of
$\sim 20$ lower than the in the non-magnetized thick shell model. In
the thin shell case, the local maximum of the thermal energy at
$1.7\times 10^{17}$\,cm (Fig.~\ref{fig:energy} upper panel) marks the
RS crossing. Afterwards, most of the energy concentrates in the
shocked external medium at distance $r > 4\times 10^{17}$\,cm when the
evolution becomes very similar to that of the unmagnetized shell.

At very early times, the Lorentz factor of the medium just behind the
forward shock is larger than that of the shell due to the initial
rarefaction. This leads to $\gamma_{\rm rel}>15$ initially until the
``transition'' radius is reached (Fig.~\ref{fig:gammas}).  The fact
that initially $\gamma_{\rm rel}>\gamma_0$ is a unique feature of
magnetized ejecta (in unmagnetized ejecta there is always
$\gamma<\gamma_0$). This initial phase appears also in the early
afterglow of the electromagnetic
model \citep{Lyutikov:2006wj,Genet:2007cl}.

The magnetization affects the (reverse) shock conditions and, as a
result, the reverse shock crosses faster a magnetized shell than an
unmagnetized one (see that the vertical blue lines appear to the left
of the vertical black lines in the lower panel of
Fig.~\ref{fig:gammas}). This feature has already been revealed in the
study of \citep{Fan:2004if}.  Our simulations show that the
rarefaction $R_2$ also crosses faster (in observer time) a magnetized
shell than a unmagnetized one.

At a timescale a few (thick shell) or several (thin shell) times that
of the duration of the burst the $\sigma=0$ and $\sigma=1$ models
display a rather similar evolution. In this stage almost all the
energy of the shell has been transferred to the shocked external medium
(Fig.~\ref{fig:energy}). The two models have the same total energy and
relax to identical asymptotic self-similar solutions (note the
similarity between the rising parts the solid thin blue and black
lines display in Fig.~\ref{fig:energy}). The forward shock emission
beyond this time cannot reveal anything about the initial
magnetization. Note, however, that there is a rather prolonged RS
crossing phase in the tail of the magnetized ejecta and some residual
energy remaining in the form of Poynting flux at later times that may
power some (energetically weak) afterglow features.

\subsection{Dissipation by the reverse shock}

Our simulations can quantitatively answer the question of how much
energy is dissipated when the reverse shock propagates into ejecta of
different magnetization $\sigma_0$ and parameter $\xi$.  For practical
reasons the simulations are limited to a few models.%
\footnote{Each simulation takes between 50 and 200 thousand hours of
    computer time using between 32 and 320 processors (depending
    whether we compute thin, thick or continuous flow models) on {\it
      Mare Nostrum}  {(\tt
      http://www.bsc.es/plantillaA.php?cat\_id=5)}. The typical
    external storage requirements of one model vary between 10 (thin)
    and 100 (thick models) gigabytes, since relatively frequent output
    of the fluid state is needed in order to obtain a
    satisfactory coverage of the fluid evolution needed for the
    post-processing calculations. }
On the other hand, they can be used to evaluate the accuracy and
limitations of previous (semi-)analytical studies
\citep{Fan:2004if,Zhang:2005ts,Giannios:2008zl} and use them as a tool
to explore a larger parameter space of $\xi$ and $\sigma_0$.

\begin{figure}
\centering \includegraphics[scale=0.45]{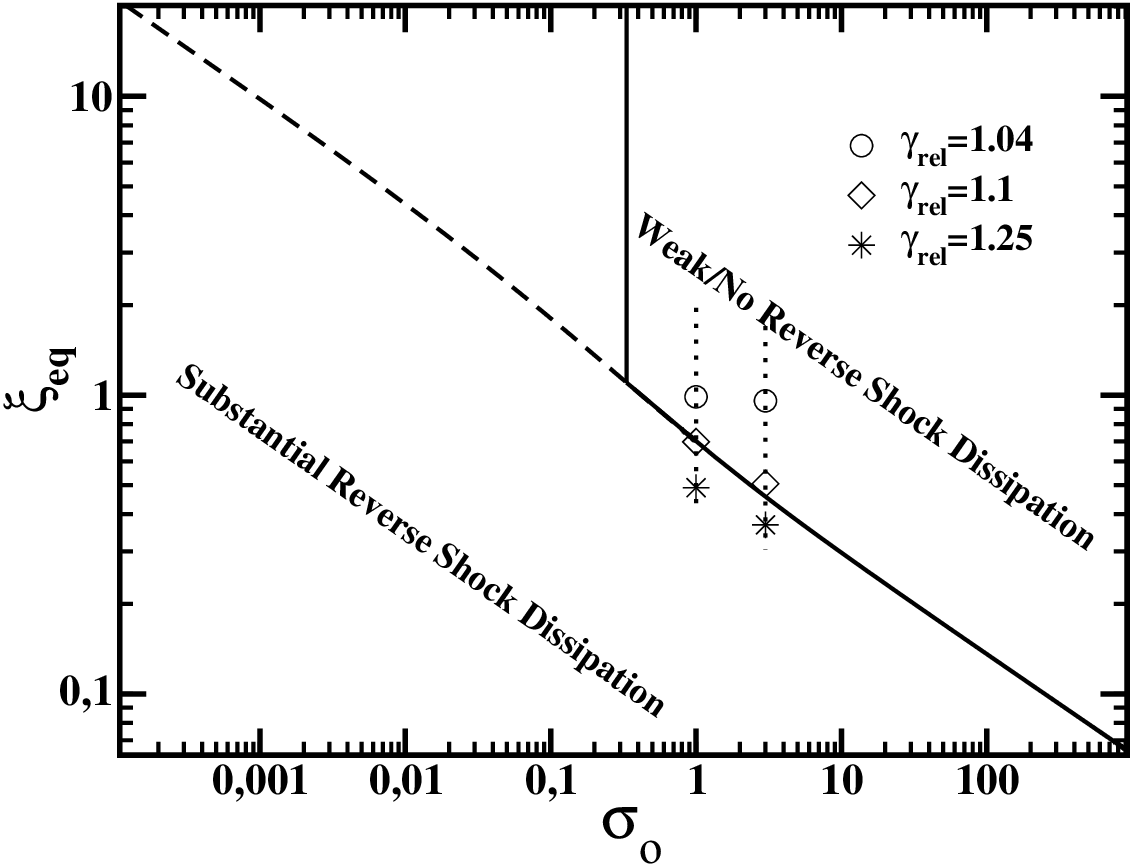}
\caption[]{Energy dissipation in the reverse shock in the
  $\xi-\sigma_0$ parameter space. The circles, diamonds and asterisks
  mark the equivalent $\xi$ of the shell for which the relative
  Lorentz factor $\gamma_{\rm rel}$ of the shocked ejecta with respect
  to the unshocked ejecta becomes 1.04, 1.1 and 1.25 respectively. The
  dotted curves show the region probed by the $\sigma_0=1$ and
  $\sigma_0=3$ simulations. The solid line marks the ``no reverse
  shock boundary'' as estimated by \citep{Giannios:2008zl}. In the
  ``weak/no reverse shock dissipation'' region, the shock converts
  less than $\sim 0.3$\% of the total energy of the shell into heat.}
\label{fig:xisig}
\end{figure}

The ``continuous flow'' models with $\sigma_0=1$ and $\sigma_0=3$
describe spherical flows of constant (as functions of radius)
luminosity $L$, magnetization $\sigma_0$ and Lorentz factor $\gamma_0$
that collide with a uniform external medium with number density
$n_{\rm e}$. Initially the interface of the two media is set at some
distance $r_0$ (see Section 3 for the choice of $r_0$) and the system
is let to evolve. With these models we can focus on the interface of
interaction between the shell and the external medium and, therefore,
we can track in great detail the formation and the strengthening of
the reverse shock with time (or equivalently radius). This kind of set
up allows us to follow the strength of the reverse shock for different
``equivalent thickness'' of shells in a single simulation.The
  idea behind the equivalent thickness is to measure the penetration
  distance from the contact discontinuity to the reverse shock and to
  assume a shell of the initial thickness $\Delta_0$ equal to this
  distance. Then we can use Eq.~(\ref{ksi}) to obtain the equivalent
  $\xi$ of the assumed shell. The consequence of this is that the more
  the reverse shock penetrates the flow, the thicker equivalent shell
  it probes for the fixed magnetization $\sigma_0$. We define
  equivalent $\xi$ as (taking into account that $E \simeq L\Delta_0/c$)
  \begin{equation}\label{eq:eqksi}
    \xi_{\rm eq} := \left(\dsfrac{3 L}{4\pi n_e m_p
        c^3}\right)^{1/6} \Delta_0^{-1/3}\gamma_0^{-4/3}
  \end{equation}
  one can see that thicker equivalent shell corresponds to lower
  $\xi$.  Effectively, a run of a continuous flow model probes a line
  of constant $\sigma_0$ in the $\xi-\sigma_0$ plane shown in
  fig.~\ref{fig:xisig}.

A convenient measure of the strength of the reverse shock is the
relative Lorentz factor $\gamma_{\rm rel}$ of the unshocked ejecta
with respect to the shocked ejecta (Eq.~\ref{eq:gamma_rel}). In
Fig.~\ref{fig:xisig}, we mark the locations where the $\gamma_{\rm
  rel}$ becomes 1.04, 1.1, 1.25 respectively.  We have argued in
\citep{Giannios:2008zl} that for $r_{\rm rs}\ll r_{\rm c}$, the
magnetization of the flow cannot prevent the formation of a shock into
the ejecta and vice versa.  The curve defined by setting $r_{\rm
  rs}=r_{\rm c}$ (thick black line in Fig.~\ref{fig:xisig}) can thus
be used as a proxy to mark the region where a reverse shock forms.

As one can see in Fig.~\ref{fig:xisig}, the region of existence of
reverse shock is larger than that estimated by
\cite{Giannios:2008zl}. The reason lies on the late RS formation that
has been revealed by the numerical simulations. This effect was
unaccounted by our previous analytic estimates. However, the analytic
prediction that the reverse shock emission from models whose
parameters are in the region $r_{\rm rs}\sim r_{\rm c}$ would not be
observed is still qualitatively valid, since the dissipation from the
shock is weak.  On the solid line $\gamma_{\rm rel}\sim 1.1$ and grows
larger only for smaller values of $\xi$.  For a $\gamma_{\rm rel}\sim
1.1$, the shock converts only a fraction of $\sim f_{\rm
  b}(\gamma_{\rm rel}-1)/(1+\sigma_0)\ll 0.1$ of the energy of the
shell into heat. Here $f_{\rm b}\sim 0.3$ for $\sigma_0\sim 1$
\citep[see e.g.,][]{Zhang:2005ts}.  Integrating the thermal energy in
the shocked shell for the snapshot for which $\gamma_{\rm rel}=1.1$,
we find that it accounts only for $\sim 0.3$\% and $\sim 0.1$\% of the
total energy of the shocked shell in the $\sigma_0=1$ and $\sigma_0=3$
runs respectively. This reveals that the shock, though there, still
dissipates very weakly ``above'' the solid line of
Fig.~\ref{fig:xisig}.

\subsection{Rescaling of the results to arbitrarily high $\gamma_0$}
\label{sec:rescaling}

Our simulations correspond to an ultrarelativistic shell of material
interacting with the external medium that has the qualitative
characteristics expected at the onset of the afterglow phase of the
GRB ejecta. Nevertheless, they have two parameters that are
unrealistic with respect to what is expected in a GRB; namely the
initial bulk Lorentz factor $\gamma_0$ and the thickness of the shell
$\Delta_0$.  GRBs are believed to originate from flows with
$\gamma_0\simmore 100$, while the thickness of the flow is likely
connected to the observed duration of the burst through $\Delta_0\sim
c t_{\rm GRB}\simless 10^{13}$ cm.  However, numerical reasons forced
us to simulate shells which have $\gamma_0\sim 15$ and $\Delta_0\sim
10^{15}-10^{16}$ cm. In this section we demonstrate a method with
which our results can reliably be rescaled to GRB-relevant parameters.

\subsubsection{Motivation for the proposed rescaling}  

We first focus on unmagnetized GRB ejecta and then extend our
discussion to include magnetization. For $\sigma_0=0$, the problem of
the interaction of an ultrarelativistic and thin cold shell with an
external medium is defined by four parameters $E$, $\Delta_0$,
$\gamma_0$ and $n_{\rm e}$. The evolution of such configuration does
not depend on the individual parameters but on a specific combination
of them that can be expressed as $\xi\propto (E/n_{\rm
  e})^{1/6}/\Delta_0^{1/2}\gamma_0^{4/3}$.  The parameter $\xi$
determines, for example, the relative Lorentz factor $\gamma_{\rm
  rel}$ of the reverse shock \citep{Sari:1995oq}. For the
ultrarelativistic flow under consideration $\gamma_{\rm rel}\simeq
F(\gamma_{\rm sh}/\gamma_0)$ (see Eq.~\ref{eq:gamma_rel}), where
$\gamma_{\rm sh}$ stands for the Lorentz factor of the shocked ejecta
when the shock reaches their rear part. This means that, for fixed
$\xi$, $\gamma_{\rm sh}$ is a fixed fraction of $\gamma_0$ {\it
  independently} of the value of $\gamma_0$. For example,
since in our $(\sigma_0, \xi)=(0,0.5)$ model we have found
that $\gamma_{\rm sh}\simeq 0.35 \gamma_0\simeq 5.2$, one can predict
that a non-magnetized shell with $\xi=0.5$ and arbitrary $\gamma_0\gg
1$ is characterized by $\gamma_{\rm sh}\simeq 0.35 \gamma_0$ at the
moment of the reverse-shock crossing.

The idea of rescaling is to solve (numerically) for the evolution of a
shell with $\gamma_{0,1}$ and $\Delta_{0,1}$, and then predict
(without simulating) the evolution of a second shell with
$\gamma_{0,2}$ (usually larger than $\gamma_{0, 1}$) and
$\Delta_{0,2}=\Delta_{0,1}(\gamma_{0,2}/\gamma_{0,1})^{-8/3}$ which
has the same $\xi$.%
\footnote{For the simplicity of the discussion we fix $E$ and $n_{\rm
    e}$ of the two shells. We generalize our expressions to arbitrary
  $E$ and $n_{\rm e}$ in the Appendix B.}
The distance from the central engine at which the reverse
shock will cross the second shell is $r_{\rm rs,2}=r_{\rm
  rs,1}(\gamma_{0,1}/\gamma_{0,2})^{2/3}$ (see Eq.~\ref{rsB}). The
same relation connects the distances $r_{\gamma_0} \simeq l
  \gamma_0^{-2/3}$ (see \citealp{Sari:1995oq}) where the shells ``1''
and ``2'' enter the deceleration phase after accumulating mass
  $M_0/\gamma_0$ of their own initial mass $M_0$. This indicates that
the characteristic distances of the shell-medium interaction for the
shell ``2'' are shifted by a factor
$(\gamma_{0,1}/\gamma_{0,2})^{2/3}$ with respect to those of the shell
``1''.  We postulate that the same is true {\it throughout} the
evolution of the system. More precisely, we claim that
rescaling the Lorentz factor from $\gamma_{0,1}\to \gamma_{0,2}$ one
can predict the evolution of a shell ``2'' by using that of the shell
``1'' providing that one also rescales the distance to $r'\to
r(\gamma'_0/\gamma_0)^{2/3}$.

We further extend the previous postulate by  adding up the
  possibility that the shell was magnetized, i.e., we extend the
  previous claim to the case $\sigma_0 \ge 0$. {\it The
    evolution of a thin, ultrarelativistic shell with
  thickness 
 \be \label{eq:Delta0}
    \Delta_{0,2}=\Delta_{0,1}(\gamma_{0,2}/\gamma_{0,1})^{-8/3} 
\ee 
is self similar to that of a shell of the same $\xi$ and $\sigma_0$
and width $\Delta_{0,1}$.}

\subsubsection{Mathematical description of the rescaling}  

\begin{figure}
\centering \includegraphics[scale=0.32]{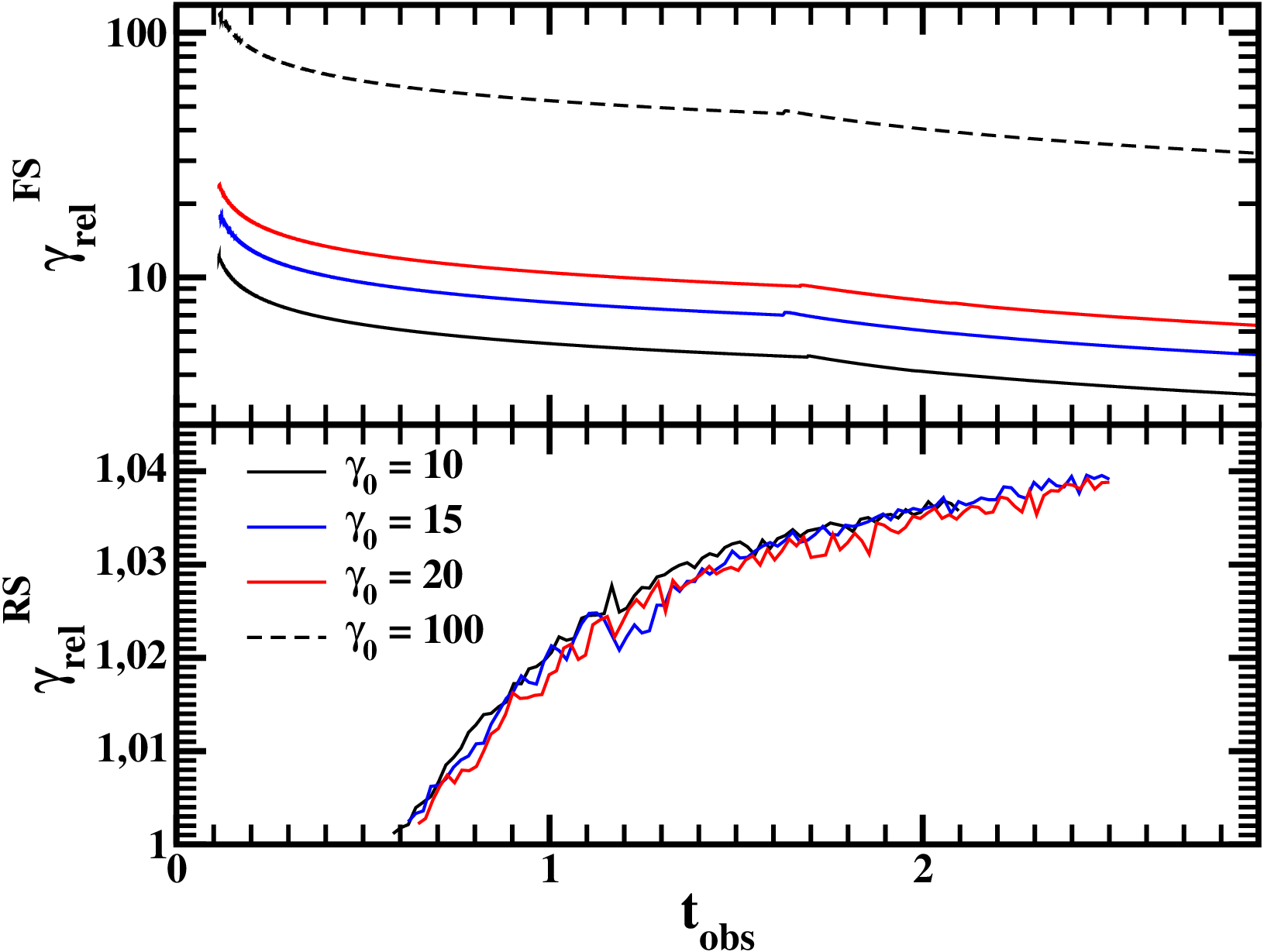}
\caption[]{Similar to Fig.~\ref{fig:gammas}, but for the models used
  in the test of the rescaling hypothesis. Black, blue and red lines
  in the upper (lower) panel show the relative Lorentz factor of the
  fluid at the FS (RS) as a function of the normalized time of
  observation for models with $\gamma_0 = 10,\ 15$ and $20$,
  respectively. As expected, the relative Lorentz factor at the front
  shock scales with $\gamma_0$ while at the reverse shock they
  coincide. The dashed line in the upper panel shows the results of a
  model obtained by rescaling to $\gamma_0=100$ the model with
  $\gamma_0=15$.}
\label{fig:scal_gamma}
\end{figure}

Here, we provide the expressions for a transformation of the solutions
for models with the same total energy $E$ and density of the external
medium $n_{\rm e}$. The more general transformation allowing for a
change of $E$ and/or $n_e$ between two models is given in the Appendix
B.  On more mathematical terms, the postulated recipe for making the
transformation from one solution to the other is the
following. Suppose the bulk Lorentz factor $\gamma_1(r_1,t_1)$ of the
shell ``1'' at is known ($t_1$ is the time in the rest frame of the
GRB engine or laboratory frame), and we define the quantity
$f:=\gamma_{0,2}/\gamma_{0,1}$. We further assume that the
bulk Lorentz factors of both shells at any other time different from
the initial one are linked by
\be\label{eq:scaling1} 
\gamma_2(r_{2},t_{2})=f\gamma_1(r_{1},t_{1}), 
\ee
where
\begin{equation}\label{eq:scaling2}
  \begin{array}{rcl}
    t_2 &=& f^{-2/3} t_1\\
    r_2 &=& r_{FS,1}(t_1)f^{-2/3} + (r_1 - r_{FS,1}(t_1))\ f^{-8/3}.
  \end{array}
\end{equation}   
Here $r_{FS,1}(t_1)$ ($r_{FS,2}(t_2)$) stands for the radius of the
forward shock of the shell ``1'' (``2'') as function of time. The
other physical quantities in the shell ``2'' can be derived from
$\gamma_2$ and using standard expressions for the forward shock
conditions. The (gas or  magnetic) pressure $P$ in the shell
and the shocked external medium, the density in the shell 
  $\rho_{\rm shell}$ and the density in the shocked external medium
$\rho_{\rm ext}$ are given by
 
\begin{equation}\label{eq:scaling3}
  \begin{array}{rcl}
    P_{2}(r_2,t_2) &=& f^{2}  P_{1}(r_1,t_1)\\
    \rho_{\rm shell,2}(r_2,t_2) &=& f^{2} \rho_{\rm shell,1}(r_1,t_1)\\
    \rho_{\rm ext,2}(r_2,t_2) &=& f \rho_{\rm ext,1}(r_1,t_1).\label{Prho}
  \end{array}
\end{equation}
There are several interesting properties of the proposed
recipe. First, the magnetization of both ejecta is the same
($\sigma_{0,1}=\sigma_{0,2}$), not only at the initial time (which
holds from our postulate), but also throughout the evolution, since
$\sigma_2(r_2,t_2) \propto P_{B,2}/\rho_{\rm shell,2}\propto
P_{B,1}/\rho_{\rm shell,1}\propto \sigma_1(r_1,t_1)$.  Second, the
$\gamma_{\rm rel}$ and its time evolution are identical for all models
(Fig.~\ref{fig:scal_gamma} upper panel), and third, the time
evolutions of $\gamma_{\rm FS}$ are just linearly shifted with
$\gamma_0$ (Fig.~\ref{fig:scal_gamma} lower panel).

\subsubsection{Invariance of the time of observation}
\label{sec:invariance}
 
An important byproduct of the transformations provided in
Eqs.~\ref{eq:scaling1}~-~\ref{eq:scaling2} is that the time of
observation defined in Eq.~\ref{eq:tobs} is invariant. The only
assumption we make is that the motion of both shells ``1'' and ``2''
is ultrarelativistic, so that the position of the FS can, generally,
be written as
\begin{equation}\label{eq:fspos}
r_{FS}(t) = c\int_0^t{\mathrm d}\tau\ \beta_{FS}(\tau) \approx ct -
(c/2) \int_0^{t}{\mathrm d}\tau \gamma_{FS}^{-2}(\tau)\, .
\end{equation}

We first demonstrate the invariance of $t_{\rm obs}$ for the
FS. Inserting Eq.~\ref{eq:fspos} into Eq.~\ref{eq:tobs} we get for the
shell ``2''
\begin{equation}\label{eq:fs2}
t_{\rm obs, FS, 2} = (c/2)\Delta_{0,2}^{-1}\int_0^{t_2}{\mathrm d}\tau_2
\gamma_{FS, 2}^{-2}(\tau_2)\, .
\end{equation}
We transform the integral as
\[
\begin{array}{rcl}
 \displaystyle{\int_0^{t_2}{\mathrm d}\tau_2 \gamma_{FS, 2}^{-2}(\tau_2)} &=& 
f^{-8/3}\displaystyle{\int_0^{t_1}{\mathrm d}\tau_1 \gamma_{FS, 1}^{-2}(\tau_1)},
\end{array}
\]
\noindent

and insert it into Eq.~\ref{eq:fs2}. After  transforming
$\Delta_{0,2}=f^{-8/3}\Delta_{0,1}$ we finally obtain the desired result
\begin{equation}
t_{\rm obs, FS, 2} =
(c/2)f^{8/3}\Delta_{0,1}f^{-8/3}\int_0^{t_1}{\mathrm d}\tau_1
\gamma_{FS, 1}^{-2}(\tau_1) = t_{\rm obs, FS, 1}\, .
\end{equation}

For a point inside the shell ``2'', different from the FS, we
have
\begin{equation}\label{eq:tobsin}
t_{\rm obs, 2} = (c/2)\Delta_{0,2}^{-1}\int_0^{t_2}{\mathrm d}\tau_2
\gamma_{FS, 2}^{-2}(\tau_2) - c\Delta_{0,2}^{-1}(r_2 - r_{FS,
  2}(t_2))\, 
\end{equation}
We see that the second term on the right hand side is also invariant
to the scaling, since $(r_2 - r_{FS, 2}(t_2))$ and $\Delta_{0,2}$ both
scale with $f^{-8/3}$. This completes our proof of the invariance of
$t_{\rm obs}$ under  the scaling relations
  Eqs.~\ref{eq:scaling1}~-~\ref{eq:scaling2}.

\subsubsection{Verification with test runs}
\label{sec:verification}

\begin{figure}
\centering \includegraphics[scale=0.32]{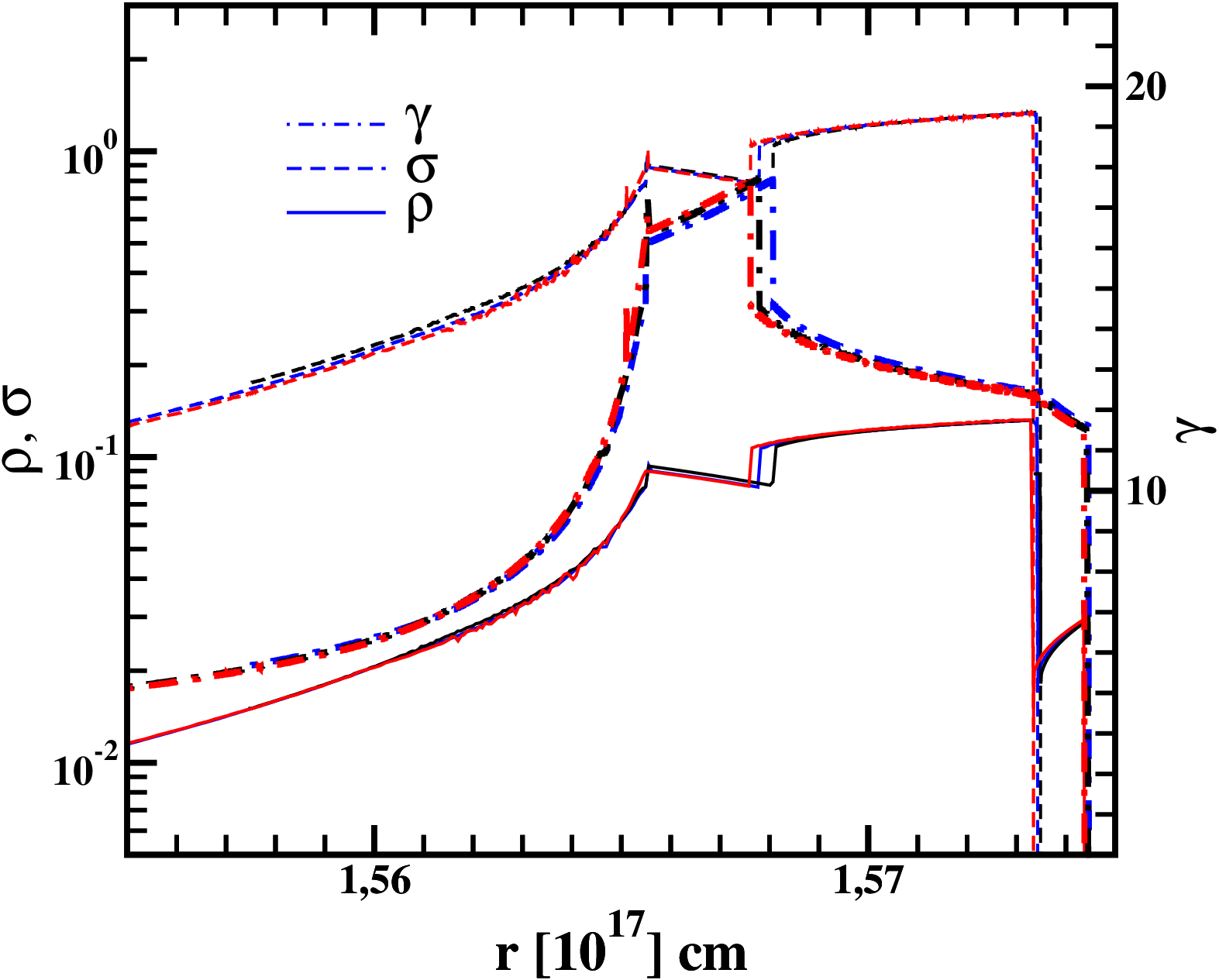}
\caption[]{Results of a test of the rescaling with $\gamma_0$. Red,
  black and blue lines show the results for  models with
  $\gamma_0 = 10,\ 15$ and $20$, respectively. Full, dashed and
  dot-dashed lines show the rest-mass density, $\sigma$ and the fluid
  Lorentz factor, respectively. Models with $\gamma_0 = 10$ and $20$
  have been rescaled using
  Eqs~(\ref{eq:scaling1})~-~(\ref{eq:scaling3}).  After
    rescaling of the results, the profiles of all models almost
    overlap. Only around the RS there are large discrepancies. The
    reason for them is that we have a finite time resolution and,
    therefore, we have to rescale our models using the closest
    discrete time we have to the one requested by the transformation
    expressed in Eqs.~\ref{eq:scaling2}.}
\label{fig:scalesnap}
\end{figure}

We have tested  numerically the postulate stated in the
  previous section for both unmagnetized and magnetized flows and
found that it is correct within a few percent accuracy.
In the following, we present three numerical models
  which share a common magnetization $\sigma_0=1$ and $\xi=1.1$. The
  rest of the parameters are: 1) $\gamma_{0,1}=10$ and $\Delta_{0,1}=
2.95\times 10^{15}$ cm, 2) $\gamma_{0,2}=15$ and $\Delta_{0,2}=1\times
10^{15}$ cm and iii) 3) $\gamma_{0,3}=20$ and $\Delta_{0,3}=2.64\times
10^{14}$ cm. We use the scaling relations given in
Eqs.~(\ref{eq:scaling1})~-~(\ref{eq:scaling3}) to conform models 1 and
3 with the model 2. On Fig.~\ref{fig:scalesnap} we show the
density, the magnetization and the Lorentz factor after
applying the scaling laws
Eqs.\ref{eq:scaling1}~-~\ref{eq:scaling3} to the models.
We show the scaling after the forward shock of the model 2 has reached
a distance $\approx 1.75\times 10^{17}$ cm. As we can see, Lorentz
factor and the magnetization $\sigma$ scale as expected. The rest-mass
density within the shell and the pressure (not shown here) are also
following Eqs.~\ref{eq:scaling3}. In Fig.~\ref{fig:scalesnap} we note
that there is a factor of 2 difference between the rescaled results
around the RS. The reason for this discrepancy arises from the finite
time resolution of our models. In order to rescale them we have to use
the closest discrete time we have available to the one requested by
the transformation expressed in Eqs.~\ref{eq:scaling2}.

\subsection{Extrapolation to $\gamma_0\simmore 100$ and ``bolometric''
  light curves}
\label{sec:lcs}

The MHD calculations presented here do not suffice to calculate the
expected emission in detail. Such a calculation of the afterglow
emission in different observed bands needs additional assumptions
related to the, poorly known, shock microphysics. These include, for
example, the energy distribution of the accelerated electrons and the
generation of magnetic fields in the shock front. Furthermore, one
needs to include radiative mechanisms such as synchrotron and inverse
Compton and allow for adiabatic losses. This detailed calculation
falls beyond the scope of this work. Instead, we apply a simple method
to get a rough estimate of the bolometric emission expected from the
different models. In order to make predictions about the properties of
the afterglow light curves associated to our numerical models, we
extrapolate them to the conventionally accepted parameter regime where
GRBs shall take place. To do this, we apply the rescaling described in
the previous Section.

We assume that in both reverse and forward shocks a fraction
$\epsilon_e \approx 0.1$ of the dissipated energy goes into
high-energy electrons, and that the electrons are in the fast cooling
regime (as is usually the case during the initial afterglow
phases). Then the resulting total emission is given by the rate at
which the shocks heat the external medium and the shell.

 From the shock conditions at the FS \citep{Sari:1995oq} we
  find that the thermal energy (in the laboratory frame) dissipated by
  the FS when it moves from $r$ to $r + \Delta r$ is
\begin{equation}
  \Delta E_{\rm th}(r, t) = 16\pi r^2\Delta r \gamma^2 n_e m_p c^2\, ,
\end{equation}
where we assume that $\Delta r\ll r$, and that $\gamma$ is
approximately constant between $r$ and $r + \Delta r$.  We assume the
luminosity of the FS to be a fraction of the thermal energy dissipated
due to the heating of the external medium by the shell in a time
interval $\Delta t = \Delta t_{\rm obs} \Delta_0/c$, where $\Delta
t_{\rm obs}$ is the difference of normalized observational time
(Eq.~\ref{eq:tobs}) between the moments in which the FS moves from $r$
to $r + \Delta r$. According to this definition, the bolometric
luminosity for the front shock reads
\begin{equation}
  \label{eq:LFS}
  L_{\rm FS} :=  16\pi n_e m_p c^2 \epsilon_e \gamma^2 r^2
  \dsfrac{\Delta r}{\Delta t}\, .
\end{equation}

For the reverse shock we estimate its luminosity assuming that
  a fraction $\epsilon_e$ of the increase of thermal energy in the
  shocked shell, as it moves from $r$ to $r + \Delta r$, is radiated
  by the RS. Thus, we define
\begin{equation}
  \label{eq:LRS}
  L_{\rm RS} : =(\Delta t)^{-1}\ max\left[0, \epsilon_e \int_{\rm shell\ r+\Delta r}4 \gamma^2
    p - \int_{\rm shell\ r}4 \gamma^2 p\right]\, ,
\end{equation}
where the $\int_{\rm shell\ r}$ denotes the integral over the volume
of the shell when the FS is at $r$. The integrand $4\gamma^2 p$ can
easily be derived from the equations of RMHD assuming an adiabatic
index $4/3$ \citep[see e.g.,][]{Mimica:2007db}.

The luminosities $L_{\rm FS}$ and $L_{\rm RS}$ can be
  normalized to the initial shell luminosity defined as 
\begin{equation}
L_0 := 4\pi r_0^2 \gamma_0^2 \rho_{\rm shell} c^3.
\end{equation}
In this way we obtain the dimensionless luminosities $l_{\rm
  FS}:=L_{\rm FS} / L_0$ and $l_{\rm RS}:=L_{\rm RS}/L_0$. 

\begin{figure}
\centering \includegraphics[scale=0.32]{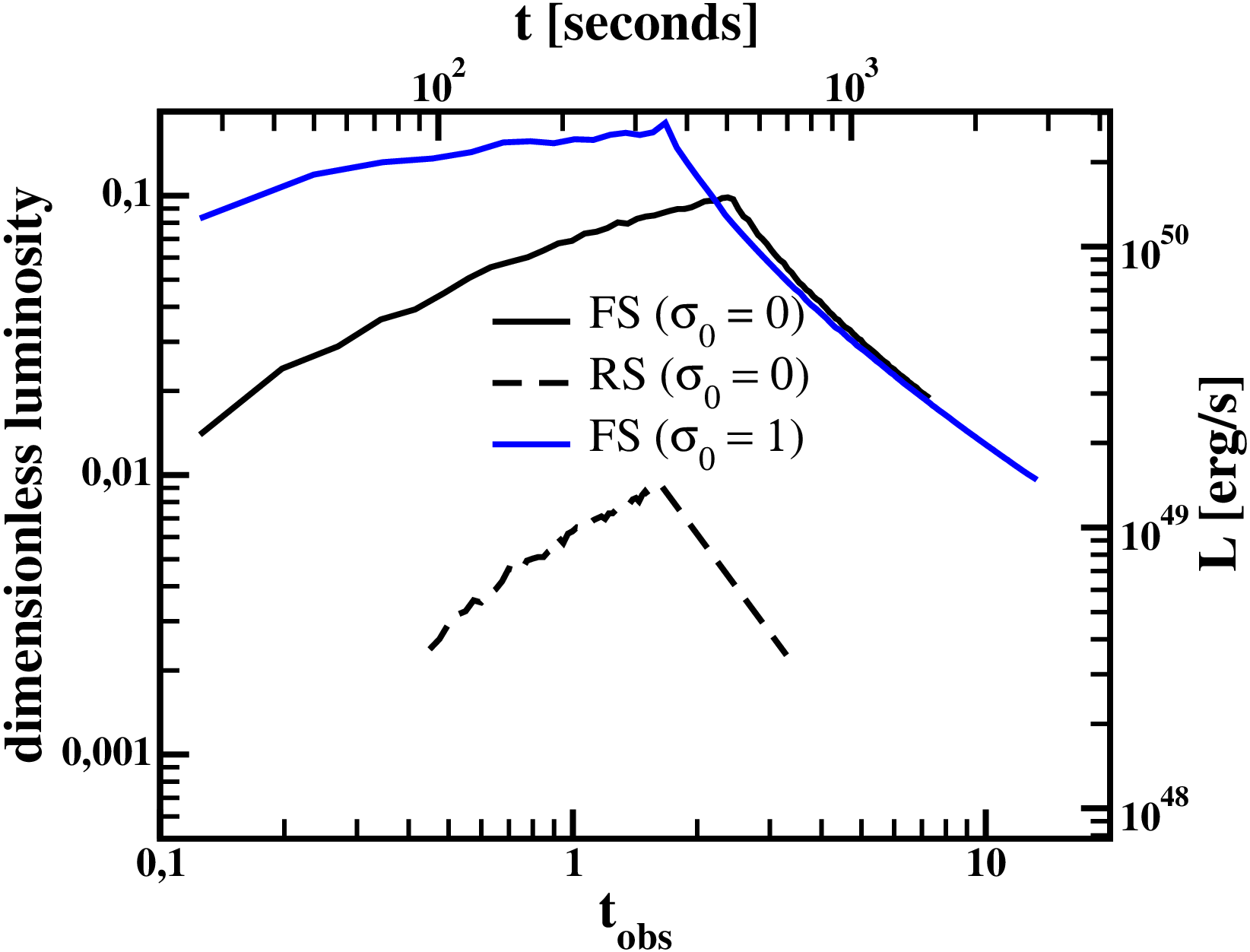}
\caption[]{Light curves for non-magnetized (black) and magnetized
  (blue line) thin shell models after scaling the $\gamma_0=15$ model
  to $\gamma_0 = 100$. The luminosity of the FS is shown in thick
  lines, while the black dashed line shows the luminosity of the
  reverse shock of the un-magnetized shell. The RS of the magnetized
  shell has a very weak dissipation and its light curve is not shown
  on the scale of the plot. Left and bottom axes show the
  dimensionless luminosity ($l_{\rm FS}$ and $l_{\rm RS}$) and the
  time of observation, respectively. The displayed luminosity profile
  can be applied to any other shell with a different $\gamma_0$
  provided that $\xi = 1.1$ (see text). Right and top axes show
  respectively the luminosity and the normalized time of observation
  for the particular case $\gamma_0 = 100$.}
\label{fig:lcthin}
\end{figure}

\begin{figure}
\centering \includegraphics[scale=0.32]{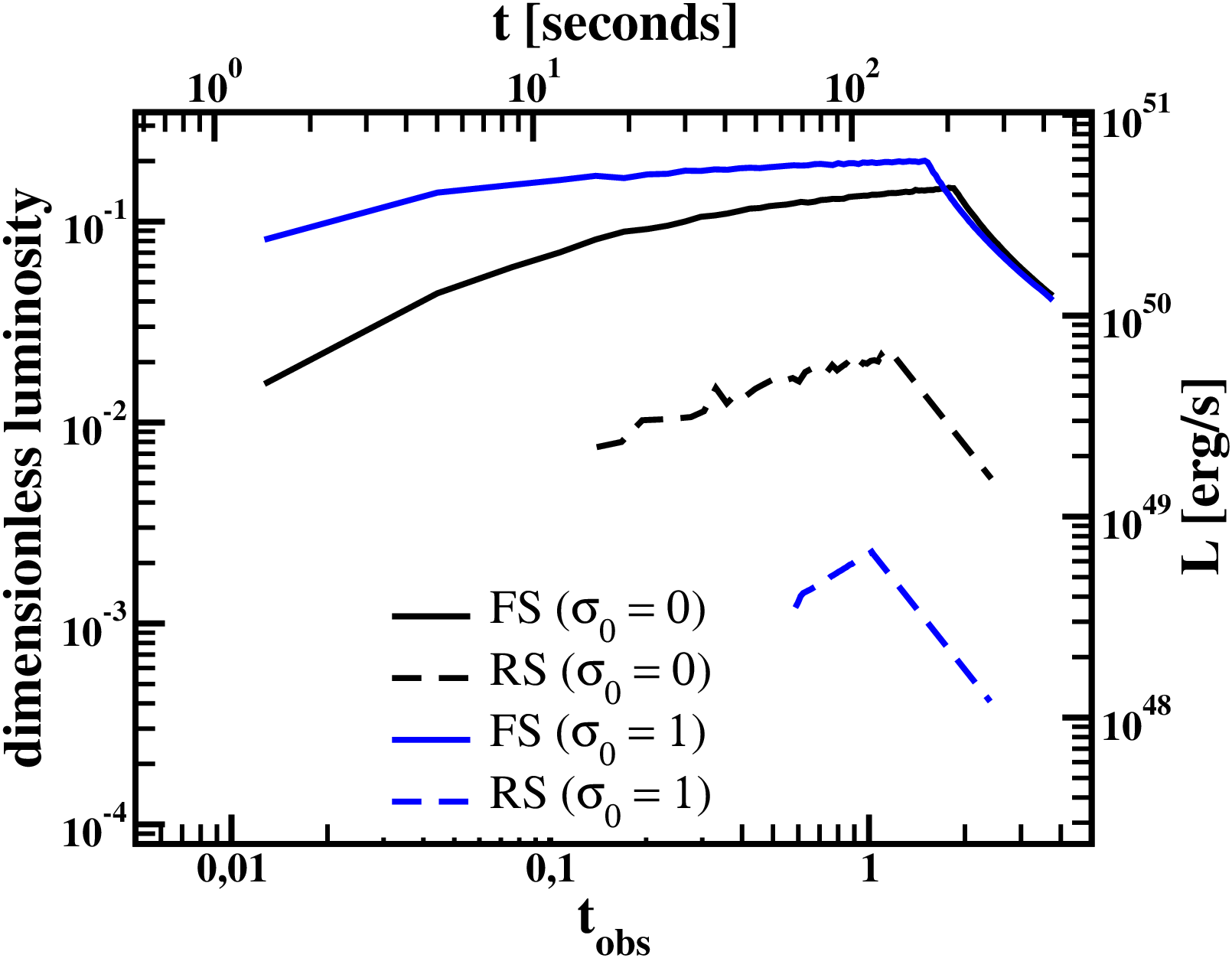}
\caption[]{Same as Fig.~\ref{fig:lcthin}, but for  a thick
  shell model ($\xi=0.5$) at $\gamma_0 = 300$. The reverse
  shock emission from the magnetized thick shell (dashed blue line) is
  much stronger than in the magnetized thin shell.}
\label{fig:lcthick}
\end{figure}

We have checked
that the normalized and conveniently scaled light curves for the test
models from Sec.~\ref{sec:verification} coincide to within a few
percent deviation. This means that we can use such a normalized light
curves to predict what would be the bolometric luminosity observed
from a shell which has, e.g., $\gamma_0 = 100$, and the same value
$\xi = 1.1$ and $0.5$ as our thin and thick shell models,
respectively. The light curves computed for the reference models with
$\gamma_0=15$ and scaled to ejecta with Lorentz factor $\gamma_0 =
100$ (thin shell; Fig.~\ref{fig:lcthin}) and $\gamma_0 = 300$ (thick
model shell; Fig.~\ref{fig:lcthick}) show a fundamental difference
between thin and thick magnetized shells. The luminosity of the RS of
magnetized thin shells is much smaller than the luminosity of the
corresponding RS in the hydrodynamic case (note that in
Fig.~\ref{fig:lcthin} the light curve of the magnetized RS does not
even show up at the scale we are considering). For thick shells, the
luminosity of the RS shock is about 10 times smaller than that of the
corresponding thick shell with $\sigma_0=0$. Hence, the detection of
the RS will be, in general, much more difficult if the shell ejecta is
magnetized than if it is unmagnetized. Indeed, if the magnetized
ejecta is thin, it is very likely that the RS is not detected at all.

\section{Discussion/Comparison with previous work}
\label{discussion}

GRB outflows may be launched by strong fields resulting in a
Poynting-flux-dominated  wind. In this case the flow can
remain strongly magnetized throughout the acceleration, collimation
and GRB emission phases all the way to the onset of the
afterglow. This is in sharp contrast to the expectations from a flow
initially dominated by thermal energy (fireball) in which magnetic
fields are dynamically unimportant. Magnetization is expected to
affect the initial phase of interaction of the GRB ejecta with the
external medium.  Since early afterglow observations are now possible
for many bursts, it is becomes topical to study the effect of
magnetization of the ejecta in the early afterglow in more detail.

Here we perform ultra-high resolution 1D RMHD simulations of the
interaction of a radially expanding magnetized ejecta with the
interstellar medium. We study the complete evolution of an
ultrarelativistic, initially uniform ejecta shell all the way to a the
self-similar asymptotic phase.  We explicitly show which is
the required resolution of numerical simulations to resolve
appropriately all the discontinuities in the flow, and to be free of
numerical artifacts.  The main findings of the work are the following:
\begin{enumerate}
\item The complete evolution can be characterized in terms of two
  parameters, namely, the Sari-Piran parameter $\xi$ and the
  magnetization $\sigma_0$. Since both parameters are defined from
  combinations of basic physical properties of the ejecta (e.g.,
  $\gamma_0$, $E$, $\Delta_0$, etc.), a single point in the
  $\{\xi,\sigma_0\}$-plane can be used to probe a variety of
  equivalent combinations of basic physical parameters. A very useful
  byproduct of this degeneracy is that one can produce numerical
  models where the shell Lorentz factor is $\gamma_0\sim$ tens and
  rescale the results to arbitrarily large $\gamma_0$. Certainly, the
  numerical difficulty of simulations with moderate Lorentz factor
  (namely, $\gamma_0 \simless 30$) is smaller than those with a large
  one (see App.~\ref{sec:resolution}). The later type of simulations
  are prone to a number of numerical artifacts that hinder an
  appropriate comprehension of the physics we are dealing
  with. The method to rescale from our reference numerical
    models to the conditions expected to occur in GRB afterglows is
    described in Sect.~\ref{sec:rescaling}, and in App.~B.

\item The amounts of energy dissipated in the reverse shock depend
  strongly on the magnetization of the flow. The reverse shock is very
  weak or absent for ejecta characterized by $\xi\simmore 0.5$
    (thin shells) and $\sigma_0\simmore 1$. The emission from the
  reverse shock is strongly suppressed for $\sigma_0\simmore 1$ except
  for high $\gamma_0\simmore 1000$ flows (or equivalently low $\xi$
  flows). More moderate values of the magnetization $\sigma\sim 0.1$
  can lead to powerful emission, in excess to that expected from
  $\sigma_0=0$ ejecta, since there is both a strong shock and a strong
  magnetic field for efficient synchrotron emission. The last
  statement assumes that there is particle acceleration
    taking place in magnetized shocks with efficiency similar to that
  of unmagnetized shocks.

\item Magnetically dominated ($\sigma_0\simmore 1$) ejecta are
  characterized by an initial rarefaction originated at the leading
  radial edge of the shell that results in a Lorentz factor of the
  front shock $\gamma>\gamma_0$. The energy that is dissipated by the
  forward shock and the expected emission at the onset of the
  afterglow is much higher for $\sigma_0\simmore 1$ ejecta with
  respect to weakly magnetized ones. ``Bolometric'' light curves are
  presented in Sec.~\ref{sec:lcs}

\item The magnetic energy of the shell is increased due to shock
  compression during the reverse shock crossing in agreement to the
  findings of \citet{Zhang:2005ts}.  We have found that the bulk of
  the magnetic energy is transferred into the shocked external medium
  on a rather sort timescale (a few to several times the burst
  duration) for the $\xi=1.1$ and $\xi=0.5$ models we have simulated.
  Several light crossing times of the initial ejecta width
    suffice for the evolution of ejecta of $\sigma\sim 1$ to become
  very similar to that of the $\sigma_0=0$ simulation. At this stage
  almost all the energy has been transferred to the shocked external
  medium and the forward shock emission is practically the same
  independently of the initial magnetization of the flow. One should
  look to the onset of the afterglow to identify effects connected to
  the magnetization.
\end{enumerate}

Comparing with previous work,  \citet{Zhang:2005ts}
significantly overestimated the reverse-shock emission for $\sigma_0 \sim 1$, 
mainly because of the assumed higher value of the relative Lorentz factor
$\gamma_{\rm rel}$ (and consequently dissipation) in the reverse shock.  
The \citeauthor{Giannios:2008zl}
analytical curve on the $\{\xi,\sigma_0\}$-plane provides a good approximation 
of when there is substantial reverse shock dissipation. The conclusion of that
work that the observed paucity of optical flashes -signatures of reverse shock
emission predicted by the fireball model- may be understood by 
$\sigma\simmore 1$ ejecta is verified by our simulations.

Modeling of the emission associated with the forward and reverse shock
can be used to compare the magnetization of the shocked external
medium to that of the shocked ejecta
\citep[e.g,][]{Zhang:2003wc}. This method has been applied in a number
of bursts
\citep{Fan:2002kx,Kumar:2003wo,Mundell:2007kx,Gomboc:2008yq}. We
caution here that such approach considers hydrodynamical shock
conditions and is therefore not reliable when the magnetization of the
ejecta is large enough to alter the dynamics of the deceleration. In
the latter case a full MHD calculation (such as the one presented
here) is needed for fruitful comparison to observations.

There are aspects of the problem of interaction of magnetized ejecta
with the external medium that have not been settled by this
work. Although we solved for the dynamical evolution of the ejecta,
the strength of the shocks as function of time and computed
approximate ``bolometric'' light curves, we did not calculate detailed
light curves in different observed bands.  For this calculation
additional assumptions related to the shock microphysics and inclusion
of radiative processes such as synchrotron and inverse Compton
emission are needed. 

Furthermore, we have not explored the high $\xi$ (or Newtonian reverse
shock) regime.  In this regime, shell spreading because of the
presence of slower and faster parts within the shell has to be
considered. For $\xi\simmore 1$, the onset of the afterglow takes
place on a later observer time and can be used to infer physical
quantities such as the Lorentz factor of the flow $\gamma_0$
\citep{Sari:1999rz,Mundell:2007rm}. The slope of the initial rising
part and the peak of the curve depend on the external medium
density profile and probably the magnetization of the ejecta. These
features are worth to be investigated in more detail. Finally since
the initial interaction lasts longer as seen by the observer,
substantial magnetic energy remains in the shell. It is possible that
dissipation of this magnetic energy is localized active regions
results in late-time flares as proposed in \citet{Giannios:2006hb}.

\section*{Acknowledgements}
The authors thank the referee for his thoughtful comments and
suggestions for improvements of this work. PM was at the University of
Valencia with a European Union Marie Curie Incoming International
Fellowship (MEIF-CT-2005-021603).  MAA is a Ram\'on y Cajal Fellow of
the Spanish Ministry of Education and Science. PM and MAA also
acknowledge the partial support from the Spanish Ministry of Education
and Science (AYA2007-67626-C03-01, CSD2007-00050). PM thanks
Jose-Maria Mart\'{\i} and Jose-Maria Iba\~{n}ez for support and
critical discussions. DG thanks Henk Spruit for useful
discussions. The authors thankfully acknowledge the computer
resources, technical expertise and assistance provided by the
Barcelona Supercomputing Center - Centro Nacional de Supercomputación.
%\end{acknowledgements}

\bibliographystyle{aa}
\bibliography{refs.bib}

\appendix

\section{Numerical method}
\label{method}

We solve equations of RMHD in 1D spherical geometry assuming that the
fluid moves only in the radial direction. The magnetic field is purely
toroidal magnetic, i.e., the magnetic field which is perpendicular to
the direction of propagation of the fluid. The system of RMHD
equations is (with the speed of light set to be $c=1$)
\begin{equation}
  \label{eq:RMHD}
  \dsfrac{\partial {\mathbf U}}{\partial t} +
  \dsfrac{1}{r^2}\dsfrac{\partial}{\partial r}\left(r^2{\mathbf
      F}\right) = {\mathbf S}\, ,
  \label{eq:1DRMHD}
\end{equation}
where the vector of unknown or conserved variables are
\begin{equation}
  {\mathbf U} = \left(\rho \gamma,\ \rho h^* \gamma^2 v,\ \rho h^* \gamma^2 - p^* - \rho\gamma,
    \ B  \right) \ .
\end{equation}
The fluxes in Eq.~\ref{eq:1DRMHD} are
\begin{equation}
  {\mathbf F} = \left(
      \rho \gamma v,
      \rho h^* \gamma^2 v^2 + p^*, \left[\rho h^*\gamma^2 - \rho\gamma\right] v, vB
  \right) \ ,
\end{equation}
and the source terms read
\begin{equation}
  \mathbf{S} = \left(
      0, \dsfrac{2p}{r}, 0, \dsfrac{vB}{r}
  \right) \ .
\end{equation}
Here $\rho$, $p$, $\gamma$ and $B$ are the fluid rest mass density,
pressure, Lorentz factor and magnetic field in the frame of the
central engine or laboratory frame. The magnetic field $B$ is measured
in Gaussian units. The total pressure is $p^* := p + B^2/2\gamma^2$,
and, the specific enthalpy is $h^* := 1 + \hat{\gamma} p/(\hat{\gamma}
- 1)\rho + B^2/\rho\gamma^2$. In our models the fluid is assumed to be
an ideal gas with the adiabatic index $\hat{\gamma} = 4/3$. We note
that we express the components of all three-vectors in the physical,
i.e., orthonormal basis.  

\citet{Romero:2005zr} discuss the solution to the Riemann problem in
case in which the magnetic field is perpendicular to the fluid
velocity and in Cartesian geometry. They show that the ratio
$B/(\gamma\rho)$ is constant everywhere except across contact
discontinuity. An analogous expression in spherical geometry, $\sigma
\propto r^2\rho$ can be derived assuming a cold magnetized fluid. In
this case the system of equations \ref{eq:RMHD} reduces to three
equations. From the continuity and the induction equation one can
easily derive the desired relation.

We use the relativistic magnetohydrodynamic code \emph{MRGENESIS}
\citep{Mimica:2005sp,Mimica:2007db}, a high-resolution shock capturing
scheme based on \emph{GENESIS} \citep{Aloy:1999rm,Leismann:2005rz} In
our code the fluid is discretized in spherical shells (zones). We use
the PPM \citep{Colella:1984sf} scheme for the spatial interpolation of
variables within numercial zones, and a HLLC \citep{Mignone:2006rz}
approximate Riemann solver to compute numerical fluxes accross zone
boundaries. The time integration is performed using a third-order
Runge-Kutta method.

\subsection{Numerical resolution}
\label{sec:resolution}

For the simulation results to be as free as possible of numerical
artifacts a large enough resolution is needed. Of particular concern
is the minimum number of iterations from the start of the simulation
which are necessary to resolve the initial evolution of the
discontinuity that separates the shell from the external medium (that
forms a Riemann problem). The knowledge of this information is
requiered since almost all RMHD codes based on approximated Riemann
solvers develop initial transient spureous behaviours at the location
of the original discontinuity. These spureous behaviours relax with
time to the correct physical solution. Therefore, since the problem is
self-similar in Cartesian coordinates, almost independently of the
initial resolution, our numerical code recovers correctly (i.e.,
within the accuracy of our method) the physical solution. However, in
spherical symmetry the problem is not strictly self-similar. Thus, the
break up of an initial discontinuity may yield to the formation of
additional discontinuities (inside of the Riemann fan but not directly
emerging from the contact discontinuity) at later times. This is
precisely what happens in the rarefaction $R_1$ in our magnetized
models (see \S~\ref{sec:magnetized_shells}), where the RS forms. If
the formation of the shock happens very close to the location of the
contact discontinuity, the initial transient artifacts in numerical
simulations may pollute the formation of the RS and yield to a wrong
numerical solution, where, e.g., the RS does not form. The way to
diminish the hampering effect of such initial transients is to
increase the numerical resolution around the initial
discontinuity. 

To eliminate the effects of the spherical geometry on the solution of
the Riemann problem, and in view of the fact that our initial shells
start at distances $R_0 \ge 10^{16}\,$cm, where the local effects of
the spherical geometry are practically negligible, in this section we
study both exact and numerical solutions of the following Riemann
problem in planar coordinates:
\begin{itemize}
  \item {\it left state}: $\rho = 1$, $p=10^{-2}$, $\gamma_0 = 15$ and $B = 15$\, ,
  \item {\it right state}: $\rho=10^{-4}$, $p=10^{-6}$, $\gamma = 1$ and $B = 0$\ .
\end{itemize}
The solution to this Riemann problem is self-similar, and consists of
a right-going shock wave separated by a contant discontinuity from a
left-going rarefaction wave. We are interested in the time $\tau$ and
the number of iterations $N_{\rm iter}$ it takes a numerical code to
obtain a correct Lorentz factor (to an accuracy of less than a
percent) of the contact discontinuity ($\gamma_{CD} = 25.56$ for this
particular problem).

\begin{figure}
\includegraphics[scale=0.3]{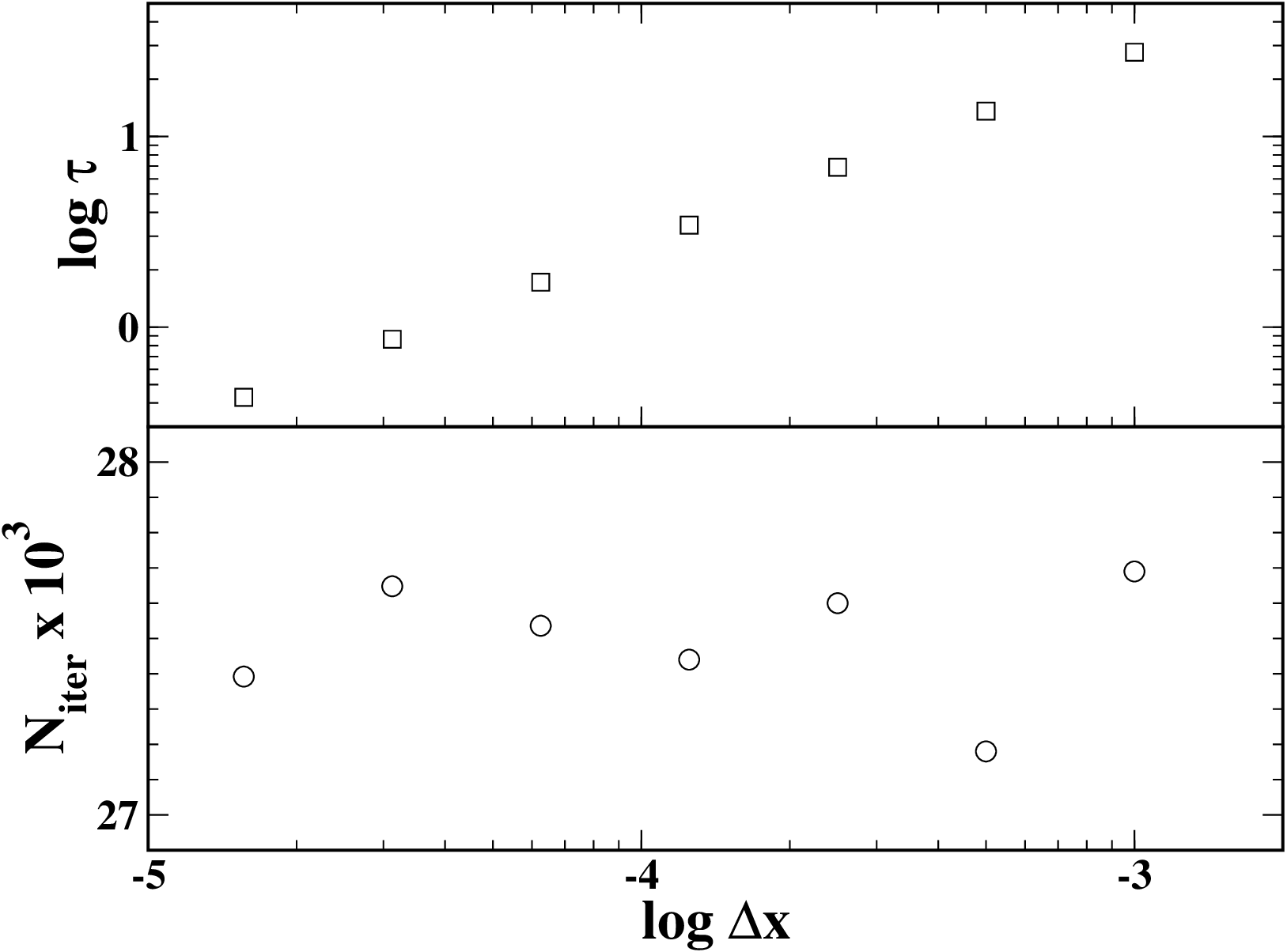}
\caption[] {Time $\tau$ (upper panel) and the number of iterations
  $N_{\rm iter}$ (lower panel) needed to resolve the Riemann problem
  in planar coordinates as a function of the spatial discretization
  $\Delta x$. }.

\label{fig:dxtau}
\end{figure}

Figure~\ref{fig:dxtau} shows the results of the test for seven
different simulations whose zone sizes have values $\Delta x =
10^{-3},\ 5\times 10^{-4},\ 2.5\times 10^{-4},\ 1.25 \times 10^{-4},
6.25\times 10^{-5},\ 3.125\times 10^{-5}$ and $1.5625\times
10^{-5}$. By fitting to the data points in the plot, we find $\tau
\propto \Delta x^{1.003\pm 0.002}$. This linear dependence can be seen
i the lower panel of Fig.~\ref{fig:dxtau}, where we see that $N_{\rm
  iter}$ is roughly independent of the resolution
(Fig.~\ref{fig:dxtau} lower panel). We find that $\tau \propto \Delta
x\ N_{\rm iter}$.

\begin{figure}
\includegraphics[scale=0.3]{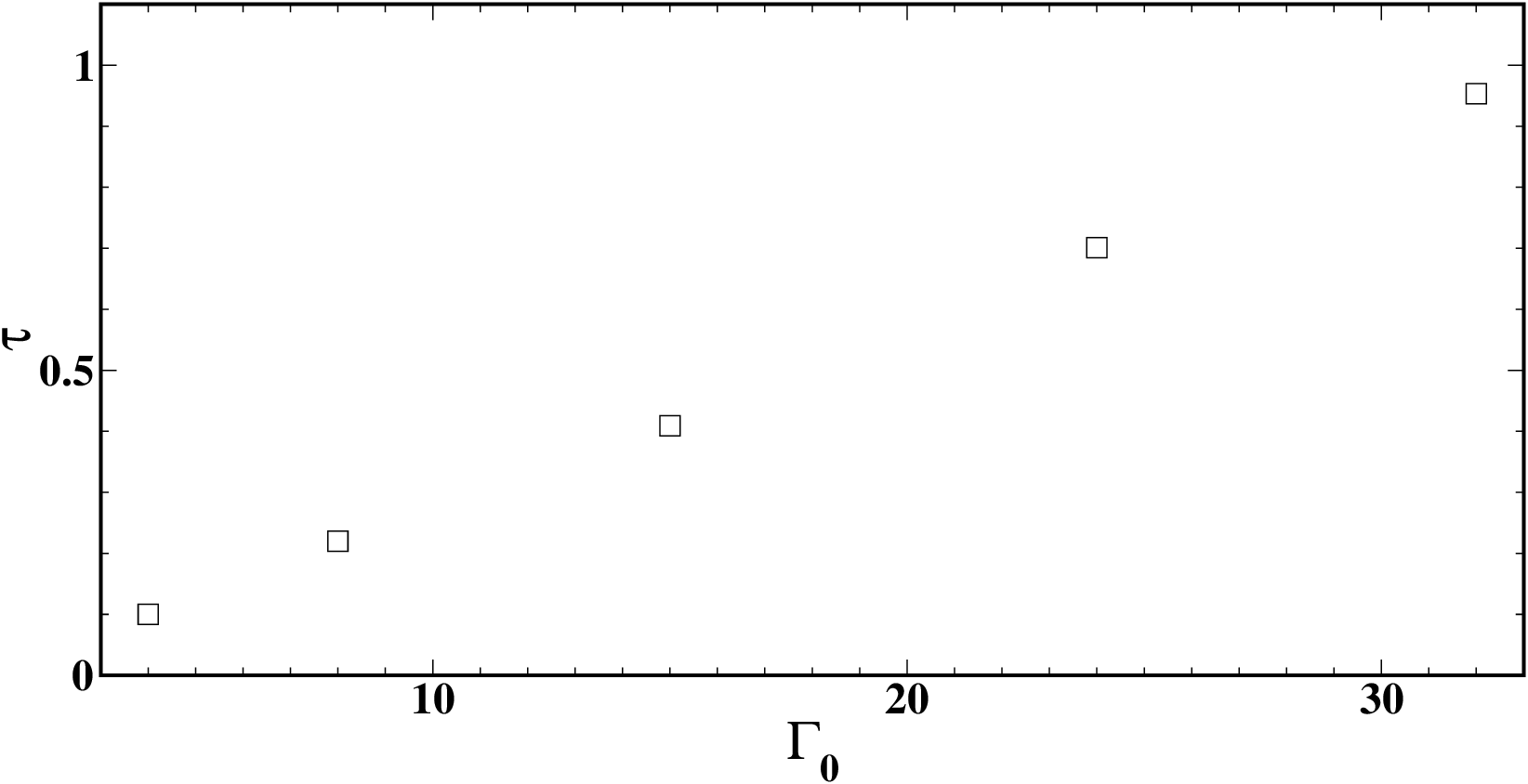}
\caption[] {Time needed to resolve the Riemann problem in planar
  coordinates as a function of the initial Lorentz factor $\Gamma_0$
  for $\Delta x = 1.5625\times 10^{-5}$. }.
\label{fig:gamtau}
\end{figure}

Of equal importance is the dependence of $\tau$ and $N_{\rm iter}$ on
the initial Lorentz factor. We modify the Riemann problem by changing
$\gamma_0$ of the left state \emph{and} the magnetic field, such that
the ratio $B/\gamma_0 = const.$ For this test we use the finest
resolution $\Delta x = 1.5625\times 10^{-5}$ for all models. Results
are shown on Fig.~\ref{fig:gamtau}. We find $\tau \propto
\gamma_0^{1.08\pm 0.02}$. Due to the constant $\Delta x$ this means
that $N_{\rm iter}$ also depends almost linearly on
$\gamma_0$. Combining results shown in Fig.~\ref{fig:dxtau} with those
of Fig.~\ref{fig:gamtau} and assuming linear dependences of $\tau$ on
$\Delta x$ and $\gamma_0$ we conclude that the required spatial
discretization for Riemann problems similar to those discussed in this
paper follows the relation
\begin{equation}\label{eq:disc}
\Delta x \propto \tau\gamma_0^{-1}\, .
\end{equation}
It has to be pointed out that the constants which are implicit in
Eq.~\ref{eq:disc} depend on the initial density and pressure ratio, as
well as the magnetization of the fluid. The result expressed by
Eq.~\ref{eq:disc} can also be interpreted in the following way: for a
fixed numerical resolution, if the Lorentz factor of the problem to be
solved grows, the time needed to relax any initial numerical pathology
also grows.

We make use of Eq.~\ref{eq:disc} to determine the maximum $\Delta x$
(or, conversely, the minimum resolution) needed to run our simulations
such that $\tau$ is much smaller than any of the characteristic
hydrodynamic time scales of our models. Particularly, $\tau$ we
warrant that $\tau$ is smaller than the time needed to form the RS in
the rarefaction fan $R_1$. 

A reduced density jump between the shell $\rho_{\rm L}$ and the
external medium $\rho_{\rm R}$ reduces drastically the numerical
complexity of the break up of the Riemann problem. Our choice of the
density jump $\rho_{\rm L}/\rho_{R} \simmore 10^4$ tries to reach the
large density contrast expected in the conditions found in GRB
afterglows (although it is still smaller than what an optimal modeling
demands). We point out that a much reduced value $\rho_{\rm
  L}/\rho_{R} \sim 10^2$ (as considered by \citealt{Mizuno:2008kx})
could be too small in regard of conditions to be met in this
astrophysical context.

\end{document}